\documentclass[10pt,a4paper]{article}
\usepackage{amssymb}
\newtheorem{thm}{Theorem}
\newtheorem{defin}{Definition}
\newtheorem{cor}{Corollary}
\newtheorem{lem}{Lemma}
\newtheorem{prop}{Proposition}

\def\C{{ \! \rm \ I\!\!\!C}}
\def\N{{ \! \rm \ I\!N}}
\def\M{{ \!  \ I\!\!M}}

\def\R{{ \! \rm \ I\!R}}

\def\E{{\!\! \rm \  I \! E}}

\def\Z{{ \! \rm Z\!\!Z}}

\def\l{{ \lambda}}

\def\tr{{\rm tr}}

\def \endsquare{ $\sqcup \!\!\!\! \sqcap$ }
\def \Ci{{C^\infty}}
\def \l {{\lambda}}
\def \e {{\epsilon}}
\def \E {{\rm I\kern-3pt E}}
\def \M {{\rm I\kern-2pt M}}
\def \S{{\rm I \kern-2pt  S}}
\def\R{{\rm I\kern-2pt R}}
\def\N{{\rm I\kern-2pt N}}
\def\Z{{\rm Z\!\!Z}}

\def\C{{\rm I\kern-5pt C}}
\def\e{{\varepsilon}}
\def \b {\bigskip \noindent}
\def \m {\medskip \noindent}
\def \Ci{{C^\infty}}
\def \s {\smallskip \noindent}
\def \l {{\lambda}}
\def \e {{\epsilon}}

\title{{ From tracial anomalies to
anomalies in \\ Quantum Field
Theory}}
\author{Alexander Cardona\footnote{and Universidad de Los Andes,
Bogot\'a, Colombia}, Catherine Ducourtioux and Sylvie Paycha }

\date{\small{\it  Laboratoire de Math\'ematiques Appliqu\'ees
\\Universit\'e Blaise Pascal (Clermont II)
\\ Complexe Universitaire des C\'ezeaux \\ 63177 Aubi\`ere Cedex, France.\\
{\rm cardona@math.univ-bpclermont.fr}\\
{\rm catherine.ducourtioux@math.univ-bpclermont.fr}\\
{\rm sylvie.paycha@math.univ-bpclermont.fr}\\}}

\begin{document}

\maketitle


\begin{abstract}
  $\zeta$-regularized traces, resp. super-traces, are defined on
a classical pseudo-differential operator $A$ by:
$${\rm tr}^Q(A):= {\rm f.p.} \, {\rm tr}(A
Q^{-z})_{\vert_{z=0}}, \quad{\rm resp.} \quad
{\rm str}^Q(A):= {\rm f.p.} \,
{\rm str}(A Q^{-z})_{\vert_{z=0}},$$   where
{\rm  f.p.} refers to the finite part and $Q$ is  an
(invertible and admissible) elliptic
reference operator  with positive order. They  are
widly used in quantum field theory in spite of
the fact that, unlike ordinary traces on matrices,
they are neither cyclic nor do they commute with
exterior differentiation, thus giving rise to
{\it  tracial anomalies}.  The purpose of this
article  is to show, on two examples,  how tracial
anomalies can lead to anomalous phenomena in
quantum field theory.
\end{abstract}

\newpage

\section*{Introduction}
In the path integral approach to quantum
field theory, $\zeta$-regularizations are used to
make sense of partition functions as
$\zeta$-determinants. Similarly,
$\zeta$-regularization procedures are used to
investigate the geometry of determinant bundles
associated to families of elliptic operators \cite{Q1},
\cite{BF}. Underlying these $\zeta$-regularizations
  is the idea of extracting a finite
part from an a priori divergent expression, such
as infinite dimensional integrals and infinite
dimensional traces.  \\ \\
Path integration  in   quantum field
theory  often gives rise to
anomalies, which we shall refer to as {\it
quantum field anomalies}.  Quantum field anomalies typically
arise from  the fact that some symmetry on the
classical level reflected in the invariance of
the classical action under some symmetry group, is
not conserved on the quantum level, namely in the
path integral built up from this classical
action.  Such anomalous phenomena  can
often be read off the geometry of determinant
bundles (see e.g.
\cite{Fr},
\cite{BF},
\cite{EM}, \cite{E}) associated to families of
operators involved in the classical action or
arising from the action of the symmetry group on
the classical action.   Here are a few
milestones of the long story of the development of
the concept of anomaly; see
\cite{Ad},\cite{BJ},
\cite{Bar}, \cite{GJ} for a perturbative approach,
see
\cite{Fu} for a path integral approach,   see
\cite{Ba},
\cite{Ber}, \cite{N} and \cite{TJZW}
 for a review.
\\
\\
On the other hand, regularized traces of the
type $\tr^Q$ (where 
$\tr^Q(\cdot):= {\rm f.p.} \tr(\cdot \,\, Q^{-z})_{\vert_{z=0}}$, 
$Q$ being the weight) give  rise to another type of anomaly,
which we refer to here as {\it tracial anomalies},
such as
\begin{itemize}
\item the coboundary $\partial \tr^Q$ of the
regularized trace $\tr^Q$
\cite{M},
\cite{MN},
\cite{CDMP},
\item the dependence on the weight
measured by $\tr^{Q_1}- \tr^{Q_2}$ where $Q_1$ and
$Q_2$ are two weights with same order
\cite{CDMP}, \cite{O},
\item the fact that it does not  commute with the
exterior  differentiation namely $[d,
\tr^Q]:= d\circ \tr^Q- \tr^Q \circ d\neq 0$ where
 $Q$ is a
family of weights parametrized by some manifold
(when this manifold is one dimensional, we use
instead the notation ${\dot \tr}^Q$)
\cite{CDMP}, \cite{P}, \cite{PR}.
\end{itemize}

Our first aim in this article,  is to show how
the use of regularized traces  and determinants in
the path integral approach to quantum field theory
can lead to    tracial anomalies, and how the
latter relate to quantum field
anomalies.  Since tracial anomalies can be
expressed in terms of Wodzicki residues
\cite{Wo}, they have some local feature which is
in turn   reflected on
the locality of anomalies in quantum field theory.
  \\ \\
 Our second aim, which is strongly linked with the
first one,  is to show how  local terms arising in
some index theorems can be
seen   as tracial anomalies; this
indirectly leads  back to some well-known relations
between anomalies in quantum field theory and
local terms in index theorems (see e.g. \cite{AG},
\cite{AGDPM}). We shall see how
\begin{enumerate}
\item the local
term in the Atiyah-Patodi-Singer theorem \cite{APS
II}
 which, for a particular family of Dirac
operators, measures a {\it phase anomaly} of a
partition function on one hand,
\item and on the other hand, the local term
in the index theorem for families from which the
 curvature on a determinant
bundle asociated to a family of Dirac
operators   can be derinved \cite{BF}, describing a
{\it (local geometric ) chiral anomaly}
\end{enumerate}
  can both be
interpreted as trace anomalies. \\
\\
In the latter case we focus  on
non gravitational anomalies, thus  restricting
ourselves to the case of a trivial  determinant
bundle. Otherwise the  curvature  arises as  a
combination of tracial anomalies and local terms
involving the underlying geometry of a fibration
of manifolds from which the determinant bundle is
built so that the
    tracial anomalies   mix  with the geometry of
the underlying fibration of manifolds  to build
geometric characteristics of the
determinant bundle such as the curvature \cite{PR}, thus
leading to a less direct relation between the two
types of anomalies, tracial and quantum field
anomalies. \\ \\
Combining the relations  we
 establish between quantum field
anomalies and tracial anomalies on one hand, local
terms in index theorems and  tracial
anomalies on the other hand, leads to the following
relations corresponding to points 1. and 2.
above:\\
\begin{enumerate} 
\item
$$
\begin{array}{ccccc}
{\begin{tabular}{|p{3.1cm}|}
\hline {\it phase} anomaly 
of \\ a partition function\\
\hline
\end{tabular}} & \Leftrightarrow &
{\begin{tabular}{|p{2.5cm}|}
\hline
   {\it tracial }anomaly\\
 $\int_0^1\dot \tr^{Q }$\\
\hline
\end{tabular}} & \Leftrightarrow &
{\begin{tabular}{|p{2.9cm}|}
\hline
local term in the \\
 {\it APS index theorem} \\
\hline
\end{tabular}}
\end{array}
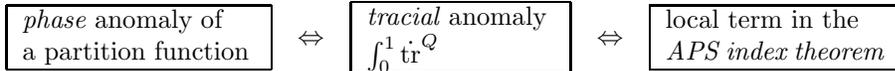
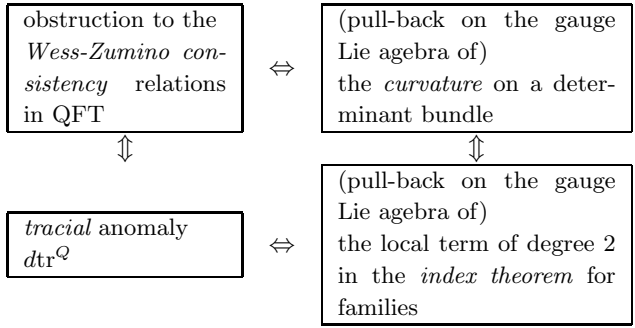
$$
\\
\item
and 
$$
\begin{array}{ccccccc}
{\begin{tabular}{|p{2.7cm}|}
\hline
{\small obstruction to the} \\
{\small {\it Wess-Zumino consistency}
relations in QFT} \\
\hline
\end{tabular}} & \Leftrightarrow &
{\begin{tabular}{|p{3.7cm}|}
\hline
{\small (pull-back on the gauge Lie  agebra of)}
\\
{\small the {\it curvature} on a determinant
bundle}
\\
\hline
\end{tabular}} \\ \Updownarrow & & \Updownarrow \\
{\begin{tabular}{|p{2.7cm}|}
\hline
{\small {\it tracial} anomaly} \\
{\small $d\tr^Q$} \\
\hline
\end{tabular}} & \Leftrightarrow &
{\begin{tabular}{|p{3.7cm}|}
\hline
{\small (pull-back on the gauge Lie
agebra of)} \\  {\small the local term of degree 2 in the
{\it index theorem} for families }\\
\hline
\end{tabular}}
\end{array}
$$
\end{enumerate}
In
particular, these relations tell us, before even
computing the various anomalies using index
theorems, that these should be local, since they
correspond to  tracial anomalies which are local
as Wodzicki residues. This approach to
anomalies seen as   Wodzicki residues is closely related in spirit to
works by J. Mickelsson and
his coworkers (see e.g. \cite{LM},  \cite{M}, \cite{MR} and very recently
\cite{AM}).
\\ \\
 The article is organized as follows. We first
recall from previous works \cite{CDMP},\cite{MN},
 \cite{P}  (section 1) how tracial anomalies occur from
taking finite parts of otherwise divergent traces.
We then briefly describe (section 2)  related
anomalies such as multiplicative anomalies (first
described in
\cite{KV}, \cite{O} and further investigated in \cite{Du}) of
$\zeta$-determinants and discuss what we call a
{\it pfaffian anomaly}, namely an obstruction
preventing  the square of the pfaffian of an
operator from coinciding with its determinant.
In section 3 we describe variations of
$\eta$-invariants as integrated tracial anomalies,
thus giving an interpretation of the local term
arising in the Atiyah-Patodi-Singer
theorem for families  \cite{APS I, APS II, APS
III} as an integrated tracial anomaly. In section
4, we discuss the geometry of determinant bundles
associated to families of elliptic operators in
relation to tracial anomalies in the spirit of
\cite{PR}, but focussing here on the case of a trivial
line bundle relevent for gauge theories.   In
section 5, we illustrate the results of section 4
by the example of families of signature operators
in dimension 3, which
give rise to a phase  anomaly
interpreted here as an integrated tracial
anomaly. It leads, via the
APS theorem,  to the well-known Chern-Simon
term in topological quantum field theory (TQFT).
In section 6, we investigate a chiral gauge anomaly
whioch can be read off the geometry of  the
determinant bundle associated to a family of
chiral Dirac operators parametrized by
connections. The pull-back on the gauge
Lie algebra of the curvature of this determinant
bundle can be interpreted as an obstruction to the
Wess-Zumino consistency relations. Here again
this obstruction  arises as a tracial anomaly. It
is a local expression given by the index theorem
for families.
 \\ Finally in Appendix A, we discuss the
relevence of the multiplicative anomaly in the
computation of the jacobian determinants
corresponding to a change of variable in a
gaussian path integral  which
underlies the computation of anomalies in
quantum field theory. We refer the reader to \cite{AM} 
for the interpretation of some gauge
anomalies in odd dimensions in terms of the  multiplicative anomaly for
what we call weighted determinants, and
\cite{CZ}, \cite{ECZ}, \cite{EFVZ}, \cite{Do} for further discussions
concerning the relevence of the multiplicative
anomaly for $\zeta$-determinants in quantum field theory.
\\ In Appendix B, following \cite{At}, \cite{Wi}, for the
sake of completeness, we briefly recall how  the
Chern-Simon term \cite{CS} in TQFT in
three dimensions (\cite{Wi}) can be derived from
the APS theorem
\cite{APS II}.
\\ \\
{\bf  Notations.}   In what follows
$M$ is a smooth closed $n$-dimensional manifold
and  $E$ a  $\Z_2$-graded vector bundle above $M$
(this includes   ordinary bundles $E$  which can be
seen as   graded bundles $E\oplus
\{0 \}$).
$Cl(M, E)$ denotes the algebra of classical
pseudo-differential operators (P.D.O.s) acting on
smooth sections of $E$ and $Ell (M, E)$, resp.
$Ell^{* } (M, E)$, resp. $Ell^{* }_{ord >0}(M,
E)$, resp.
$Ell^{*  adm}_{ord >0}(M, E)$ the set of
elliptic, resp. invertible elliptic, resp.
invertible elliptic with positive order, resp.
invertible admissible elliptic classical
pseudo-differential operators which have positive
order.  A {\it weight} is an element of $Ell^{*
adm}_{ord >0}(M, E)$ often denoted by
$Q$ and with order $q$ (in the self-adjoint case, one can drop the
invertibility condition as we
explain further along).
\section{Weighted Trace Anomalies}
Given a weight $Q$ and   $A$ in $Cl(M, E)$, the
map $z\mapsto {\rm tr}(AQ^{-z})$ is meromorphic
at $z=0$ with a pole of order $1$ and following
\cite{CDMP}  we call {\it $Q $-weighted trace of $A$},
resp. {\it $Q $-weighted super-trace of $A$}
the expression:
\begin{equation}\label{eq:trQ}
{\rm tr}^Q(A):= {\rm f.p.} \left({\rm tr} (  A
Q^{-z})\right)_{ \vert_{z= 0} }, \quad {\rm resp.} \quad {\rm str}^Q(A):=
{\rm f.p.} \left({\rm str} (
A Q^{-z})\right)_{ \vert_{z= 0} }
\end{equation}
where ${\rm f.p.}$ means we take  the finite part of the
 expansion at $z=0$ of the meromorphic function
 ${\rm tr} ( A Q^{-z})$, resp. ${\rm str} (  A
Q^{-z})$ and where $str(\cdot):= tr(\Gamma
\cdot)$, $\Gamma$ denoting the grading operator
which can be seen as a multiplication operator
acting fibrewise on the fibres of $E$.
 \\  \\
{\bf Remark.} The definition of a
complex power
$Q^{-z}$ involves a choice of spectral cut for the
admissible operator $Q$.   In order to
simplify notations we drop the explicit mention of
the spectral cut in the definition of the weighted trace. In the case
when $Q$ is 	a positive operator, any ray in $\C$ different from the
positive real half line serves
as a ray in the spectrum of the leading symbol  and an easy computation
yields
 ${\rm tr}^{Q^k}= {\rm tr}^Q$ for any  positive integer $k$.
\\
\\
 We also define the {\it Wodzicki residue } of
$A$:
$${\rm res}(A):={\rm ord Q}\cdot  {\rm
Res}_{z=0}\left( {\rm tr} (  A Q^{-z}) \right),
$$
resp.
$${\rm sres}(A):={\rm ord Q}\cdot
{\rm Res}_{z=0}\left(
{\rm str} (  A
Q^{-z}) \right)= {\rm res}( \Gamma A), $$
where the
order of the operator $Q$ is denoted by ord$Q$.
Unlike weighted traces, the Wodzicki residue does
not depend on the choice of $Q$ and defines a
trace on the algebra of classical P.D.Os. Another
important feature of the Wodzicki residue  is that
it can be described as an integral of local
expressions involving the symbol of the operator
\cite{Wo}:
\begin{equation}\label{eq:localres}
{\rm res}(A)= {1 \over (2\pi)^n} \int_M
\int_{\vert
\xi \vert=1} {\rm tr}_x\left
( \sigma_{-n} (x,
\xi)\right) d
\xi d\mu(x)
\end{equation}
where $n$ is the dimension of $M$,  $\mu$ the volume measure on $M$,
${\rm tr}_x$ the trace on the
fibre above $x$ and
$\sigma_{-n}$ the homogeneous component of order $-n$ of
symbol of the classical pseudo-differential
operator
$A$.
\\ \\
 When $Q$ has positive leading symbol, we can recover the
$\zeta$-regularized trace (\ref{eq:trQ}) using a
heat-kernel expansion. Indeed, via  a
 Mellin transformation \cite{BGV}, one can show that
(see e.g. \cite{P}):
$$ {\rm f.p.}\left( {\rm tr} (  A
Q^{-z}) \right)_{  \vert_{z= 0 }}=
{\rm f.p.}\left( {\rm tr} (  A e^{-\e Q})
\right)_{ \vert_{\e= 0} }-{
\gamma\over {\rm ord} Q }
\cdot {\rm res} (  A) $$
$$ {\rm resp.} \quad {\rm f.p.}\left(
{\rm str} (  A Q^{-z}) \right)_{  \vert_{z= 0 }}=
{\rm f.p.}\left( {\rm str} (  A e^{-\e Q})
\right)_{
\vert_{\e= 0} }-{
\gamma\over {\rm ord} Q } \cdot {\rm sres} (
A)
$$ where
$\gamma$ is the Euler constant. Thus, if
 res$(A)=0$, resp.  sres$(A)=$ res$(\Gamma A)=0$ in
the $\Z_2$-graded case, we find:
$$  {\rm tr}^Q (  A
 )  = {\rm f.p.}\left( {\rm tr}
(  A e^{-\e Q}) \right)_{ \vert_{\e= 0} } $$
$$ {\rm resp.} \quad
{\rm str}^Q (  A  )  = {\rm f.p.}\left(
{\rm str} (  A e^{-\e Q}) \right)_{
\vert_{\e= 0} }. $$  \\
   The notion of weighted
trace can be extended to the case when
$Q$ is a non injective self-adjoint elliptic
operator with positive order. Being elliptic, such
an operator has a finite dimensional kernel and the
orthogonal projection
$P_Q$ onto this kernel is a P.D.O. of finite rank. Hence, since $Q$ is an
elliptic
operator  so is the operator
$Q+P_Q$, for the ellipticity  is a  condition  on the leading symbol which
remains unchanged when adding $P_Q$. Moreover, $Q$ being self-adjoint the
range of $Q$ is given by
$R(Q)= \left(\ker Q^*\right)^\perp= \left(\ker Q \right)^\perp$ so that
$Q^\prime:=Q+P_Q$ is onto.
 $Q^\prime$ being  injective and onto is invertible and being self-adjoint,
and therefore admissible,
it  lies in
 $Ell^{* adm}_{ord >0} (M, E)$   (it has the same order as $Q$) and we can
define
${\rm tr}^{Q^\prime} (A)$, resp. ${\rm str}^{Q^\prime} (A)$.
A straightforward computation shows that:
\begin{equation}\label{eq:trQprime}
{\rm tr}^{Q^\prime} (A)=  {\rm f.p.} \left(
{\rm tr} (Ae^{-\e
Q})\right)_{\vert_{\e=0}}, \quad {\rm resp.} \quad {\rm str}^{Q^\prime}
(A)=   {\rm f.p.}
\left( {\rm str} (Ae^{-\e Q})\right)_{\vert_{\e=0}}.
\end{equation}
  We pay   a price for having left out
divergences when taking the finite part of
otherwise diverging expressions, namely the
occurence of
  {\it   weighted trace anomalies}. They will play
an important role in what follows and we shall
  show later on how they relate to  chiral
(gauge) anomalies. \\ \\
In order to describe weighted
trace anomalies, it is useful to recall
properties of logarithms of admissible elliptic
operators.  The logarithm of a classical P.D.O.
$A\in Ell^{*  adm}_{ord >0} (M, E)$ is defined by
$\log A= {d\over dz}_{\vert_{z=0}} A^z$,  and depends on
the spectral cut one chooses to define the complex
power
$A^z$.  Although the logarithm
of a classical P.D.O. is not classical, the bracket  $[\log Q, A]$ and the
difference   ${\log Q_1\over q_1}- {\log Q_2\over
q_2}$ of two such logarithms  are classical
P.D.O.s.
\subsubsection*{A first weighted trace anomaly:  the coboundary}
It is by now a well known fact that, despite their
name, weighted traces are not traces; given
$A, B\in Cl(M, E)$  we have \cite{M},\cite{MN},  \cite{CDMP}:
\begin{equation}\label{eq:coboundary}
\partial {\rm tr}^Q(A, B)
 = {\rm tr}^Q([A, B])=
-{1\over {\rm ord}Q}
{\rm res}
\left(A[\log Q,  B]\right)
\end{equation}
where
$\partial {\rm tr}^Q $ denotes the coboundary of the
linear functional ${\rm tr}^Q$ on the Lie algebra $CL(M,
E)$ in the Hochschild cohomology. This coboundary
corresponds to  the Radul cocycle  in the physics
literature \cite{R}, \cite{M}. \\ \\
This extends to weighted super-traces:
\begin{lem}\label{l:scoboun} Let $A, B\in Cl(M, E)$ be
two  P.D.O.s and let $Q$ be an even admissible
elliptic invertible operator, all
  acting on sections of some
super-vector bundle $E:= E^+\oplus E^-$. Then
\begin{equation}\label{eq:scoboun}
\partial {\rm str}^Q(A, B)
 = {\rm str}^Q(\{A, B\})
 = -{1\over {\rm ord}Q}
{\rm sres}
\left(A\{\log Q,  B\}\right).
 \end{equation}
 where $\{A, B\}:= AB +
(-1)^{\vert A\vert
\cdot
\vert B\vert}BA$ with $\vert A\vert=0$,
resp. $\vert A\vert=1$ if $A$ is even, resp. $A$
is odd.
\end{lem}

{\it Proof.} Writing
$A:=
\left[
\matrix{ A_{++}& A_{+-}\cr
A_{-+} & A_{--} \cr}\right]$,
$B:= \left[ \matrix{
B_{++}& B_{+-}\cr
B_{-+} & B_{--} \cr}\right]$, one easily sees it is sufficient to
  check the formula  for the odd operators  $  \left[ \matrix{
0&A_{+-}\cr
A_{-+} & 0 \cr}\right]$  and $  \left[ \matrix{
0&B_{+- }\cr
B_{-+} & 0 \cr}\right]$, since the result for the even part follows from
(\ref{eq:coboundary}).\\ \\
Let us therefore consider two odd operators $A=
\left[\matrix{ 0& A^-\cr A^+ &0 \cr}\right]$ and
$B= \left[\matrix{ 0& B^-\cr
B^+ &0 \cr}\right]$ acting on sections of some super-vector bundle $E:= E^+
\oplus E^-$. We have:
\begin{eqnarray*}
{\rm str}^Q (\{A, B\})&=& {\rm tr}^Q(
\Gamma \{ A, B\})\\
&=& {\rm tr}^Q( \Gamma AB+
\Gamma BA)\\
&=&{\rm tr}^Q( -A^+B^-+
B^-A^+-B^+A^- + A^- B^+)\\
&=&{\rm tr}^Q([
B^-,A^+])+ {\rm tr}^Q([A^-, B^+])\\
&=&  {1 \over {\rm ord} Q}{\rm res}(A^+[\log Q, B^-])-{1
\over {\rm ord} Q} {\rm res} (A^-[\log Q, B^+])\\
 &\quad& \textrm{where we have used (\ref{eq:coboundary})}\\ &= & -
{1 \over {\rm ord} Q} {\rm res}(\Gamma A^+
\{\log Q, B^- \})-{1 \over {\rm ord} Q}
{\rm res}(\Gamma A^-
\{\log Q, B^+\})\\
&\quad& \textrm {where we have
used the fact that Q (and hence log Q) is even}\\
 &=& -{1 \over {\rm ord} Q} {\rm sres}
 (  A \{\log Q, B\}).
\end{eqnarray*}
\endsquare 
\subsubsection*{A second weighted trace
anomaly: the dependence on the  weight}
Weighted traces depend on the choice of the weight
in the following way.  For
$Q_1, Q_2
\in Ell^{*adm}_{ord >0} (M, E)$ with  orders
$q_1, q_2$   we have \cite{CDMP}:
\begin{equation}\label{eq:trQ1Q2} {\rm tr}^{Q_1} (A)- {\rm tr}^{Q_2}
(A) =
- {\rm res }\left(   A \left({\log Q_1\over q_1}-
{\log Q_2\over q_2}\right)\right).
\end{equation}
In a similar way, for weighted supertraces we
have:
\begin{equation}\label{eq:strQ1Q2}
{\rm str}^{Q_1} (A)- {\rm str}^{Q_2}
(A) =
- {\rm sres} \left(   A \left({\log Q_1\over
q_1}- {\log Q_2\over q_2}\right)\right).
\end{equation}
This extends to variations of traces of one
parameter families of operators
$\{Q_x, x\in X\}$  in $Ell^{*
adm}_{ord >0} (M, E)$ with constant order $q$,
and common spectral cut,
$X$ being some smooth manifold.  For a  given $A\in
Cl(M, E)$ we have \cite{CDMP}, \cite{PR}, \cite{P}:
\begin{equation}\label{eq:dtr}
[ d ,{\rm tr}^{Q } ] (A):=  d{\rm tr}^Q
(A)= -{1
\over q} {\rm res}( A \, d \log Q ),
\end{equation}
and similarly for weighted supertraces:
\begin{equation}\label{eq:dstr}
[ d ,{\rm str}^{Q }]  (A):=  d{\rm str}^Q
(A)= -{1
\over q} {\rm sres}( A \, d
\log Q ).
\end{equation}
 Using the
Fr\'echet Lie group structure on the set $CL_0^*(M, E)$  of zero order
invertible P.D.O.s  to
define $e^{tB}, t\in \R$  for  a zero order P.D.O. $B$ and applying
(\ref{eq:dtr}) to
$Q_t:= e^{-t B} Q e^{ tB}$  yields:
$$\dot {\rm tr}^{Q_t}  (A):= \left[d \over
dt\right]{\rm tr}^{Q_t}  (A) =  {1 \over q} {\rm res}( A [B,
\log Q])= \partial {\rm tr}^Q(A,B), $$
so that the
anomaly (\ref{eq:coboundary}) can be seen as a manifestation of the
anomaly (\ref{eq:dtr}). A similar computation would lead us
from (\ref{eq:dstr}) to (\ref{eq:scoboun}).  Note that since the
difference of two logarithms of admissible
operators of same order is classical, so is the
differential of the logarithm of a  family of such
operators. \\
\\  An important observation in view of what
follows is that all these weighted trace
anomalies (\ref{eq:coboundary}), (\ref{eq:trQ1Q2}), (\ref{eq:dtr}),
(resp. (\ref{eq:scoboun}),
(\ref{eq:strQ1Q2}), (\ref{eq:dstr})) being Wodzicki residues (resp.
superresidues) of some operator,  can be expressed
in terms of integrals on the underlying manifold
$M$ of local expressions involving the symbols of
that operator.    \\ \\
{\bf Terminology.} Inspired by the terminology
used for anomalies in quantum field theory, we
shall refer to
 $A\mapsto [d,
{\rm tr}^Q](A)$, $A\mapsto [d,
{\rm str}^Q](A)$ and
$A\mapsto  \dot{{\rm tr}}^Q(A)$ as {\it
infinitesimal trace anomalies} and to $A\mapsto
\int_0^1
\dot{{\rm tr}}^Q(A)$
 as {\it integrated trace anomalies}. Striclty
speaking, as we shall see in the sequel,
anomalies in quantum field theory arise not so
much as  maps $[d, \tr^Q]$ but rather as their
value
$[d,
\tr^Q](A)$ for specific operators $A$; the
sign  of a Dirac operator in odd dimensions is one
example of pseudo-differential operator $A$ we
shall come across in the expression of the
phase anomaly described in section 5.
\subsubsection*{Extending weighted traces to
logarithms}
 In finite dimensions,
determinants are exponentiated  traces of
logarithms; we
  extend weighted traces to  logarithms of  pseudo-differential  operators
in order  to define
determinants in infinite dimensions. \\ \\
Given
$A, Q
\in Ell^{*adm}_{ord >0} (M, E)$   we set (see
\cite{KV}, \cite{O}, \cite{Du}, \cite{L}):
\begin{equation}\label{eq:trQlog}
{\rm tr}^Q (\log A):=  {\rm f.p.}\left( {\rm tr}
(\log A Q^{-z}) \right)_{{\vert_{z=
0}}}.
\end{equation}
As before, $Q$ is referred to as
the weight and
${\rm tr}^Q (\log  A)$ as the $Q$-weighted trace of
$\log A$. Underlying this definition, is a choice
of a determination of the logarithm which we shall
not make explicit in the notation unless it is
strictly necessary. \\ \\
{\bf Theorem \cite{O} (see also
\cite{Du})}  {\it For
$Q_1, Q_2, A
\in Ell^{*adm}_{ord >0} (M, E)$ with  orders
$q_1, q_2$ and $a$ respectively, }
\begin{eqnarray}\label{eq:thmO} {\rm tr}^{Q_1} (\log A)- {\rm tr}^{Q_2}
(\log A) \!\!\! &=& -
 {1 \over 2} {\rm res }
\left[\left(\log A-{a \over
q_1}
\log Q_1\right)
\left({\log Q_1\over q_1}- {\log Q_2\over
q_2}\right)\right] \nonumber \\
&-& \!\!\! {1\over 2  }{\rm res}
\left[ \left(\log A-{a \over
q_2}
\log Q_2\right)\left({\log Q_1\over q_1}- {\log Q_2\over
q_2}\right)\right]
\end{eqnarray}

\section{From multiplicative anomalies for
$\zeta$-determinants to Pfaffian anomalies }
 \medskip \noindent We recall here some basic properties of
$\zeta$-determinants of admissible operators.
  For an admissible elliptic  operator $A\in Ell^{adm}_{ord
>0}(M, E)$  of positive order with non zero eigenvalues,
the function
$\zeta_A(z):=
\sum_{\l\in Spec(A)}\l^{-z}$ is holomorphic at
$z=0$ and we can define the $\zeta$-determinant of
$A$:
\begin{equation}\label{eq:detzeta}
{\rm det}_\zeta (A):=\exp\left(
-\zeta_A^\prime(0)\right)=\exp  {\rm tr}^A( \log A) .
\end{equation}
 {\bf Remark.}   In fact    physicists often  consider {\it relative
determinants}
i.e. expressions of the type
$${{\rm det}^Q(A)\over {\rm det}_\zeta(Q)}= \exp {\rm tr}^Q \left(\log
A-\log Q\right)$$
combining  a weighted determinant ${\rm det}^Q(A):= \exp {\rm tr}^Q(\log
A)$ (a notion introduced
in \cite{Du}) with the
$\zeta$-determinant of a fixed reference operator (the weight $Q$ here).
Weighted and $\zeta$-determinants are related by a Wodzicki residue
$${\rm det}_\zeta(A)=
{\rm det}^Q(A) \exp \left(-{a\over 2}  {\rm res}\left( {\log Q \over
q}-{\log A \over
a}\right)^2\right).$$
   \\
    The $\zeta$-determinant is invariant under inner automorphisms of
$Ell^*_{ord >0}(M, E)$.
Indeed,  let $A$ be an operator in $  Ell^{*
adm}_{ord >0}(M, E)$ and let $C\in CL(M, E)$  be
invertible, then
$CAC^{-1}$ lies in
$ Ell^*_{ord >0}(M, E)$ and is also admissible since an inner automorphism
on P.D.Os induces an inner
automorphism on  leading symbols
$\sigma_L(CAC^{-1})=\sigma_L( C)
\sigma_L(A)
\sigma_L(C)^{-1}$ and hence leaves both the spectra  of the operator and of
its leading symbol
unchanged. Moreover, using the fact that, given $Q \in Ell^*_{ord >0}(M,
E)$ admissible, we have
$\log CAC^{-1}=
\log A$ and   ${\rm tr}^{CQC^{-1}}
(C\log AC^{-1})= {\rm tr}^Q(\log A)$, a fact which
can easily be deduced from the definition of
weighted traces (see \cite{CDMP}), it follows that:
\begin{equation}\label{eq:detcov}
\det\nolimits_\zeta (CAC^{-1})= \det\nolimits_\zeta(A).
\end{equation}
\subsubsection*{Multiplicative anomaly \cite{KV}}
Another type of anomaly which is closely related to weighted trace
anomalies is the multiplicative
anomaly of $\zeta$-determinants. The Fredholm determinant is multiplicative
but   the
$\zeta$-determinant  is not, this leading to an
  anomaly
    $F_\zeta(A, B) := { {\rm det}_\zeta (AB)
\over  {\rm det}_\zeta (A ){\rm det}_\zeta (B )}$ which
reads \cite{KV},  \cite{Du}:
\begin{eqnarray}\label{eq:MultAn}
\log F_\zeta(A, B)&=&
  {1 \over 2 a}{\rm res} \left(\left(   \log A -{a \over
a+b}
\log (AB) \right)^2\right) \nonumber \\
&+&  {1 \over 2 b}{\rm res} \left(\left(   \log B -{b
\over a+b}
\log (AB) \right)^2\right)\\
&+&  {\rm tr}^{AB} \left(
\log (AB)- \log A- \log B\right) \nonumber
\end{eqnarray}
 for any two operators $A, B \in Ell^{*
adm}_{ord >0} (M, E)$ of order
$a$  and $b$, respectively. Specializing to $B= A^*$, the adjoint of $A$
for the $L^2$ structure
induced by a Riemannian metric on $M$ and a Hermitian one on $E$, in
general we have $F_\zeta(A,
A^*)\neq 0$ and hence:
\begin{equation}\label{eq:MultAn2}
{\rm  det}_\zeta(A^*A) \neq\vert {\rm
det}_\zeta(A )\vert^2.
\end{equation}Weighted determinants are not multiplicative either and
their {\it multiplicative anomaly}  can be expressed using a
Campbell-Hausdorff formula for
P.D.O.s, see \cite{O}, \cite{Du}, see also \cite{AM} where such expressions
are used to derive gauge anomalies in quantum field theory.
\subsubsection*{$\zeta$-determinants for
self-adjoint operators}
$\zeta$-determinants  take a specific form for
self-adjoint operators, which involves the
$\eta$-invariant. \\
Let $A\in Ell^*_{ord >0} (M,
E)$ be a self-adjoint elliptic (classical)
pseudo-differential operator. The $\eta$-invariant
first introduced by Atiyah, Patodi and Singer \cite{APS
I, APS II,APS III} is defined by:
$$\eta_A(0) :=
{\rm tr}^{\vert A\vert}(  {\rm sgn}(A)),$$
where  the  classical P.D.O.  sgn$(A):= {A   \vert
A\vert^{-1}}$ can be seen as  the sign of $A$.
Since res (sgn $A ) =0$ \cite{APS I}, the
renormalized limit f.p.
$\left({\rm tr} ({\rm sgn} A \vert A \vert^{-z})\right)_{\vert_{z=0}}$ is
in fact an ordinary
limit so that
$\eta_A(0)= \lim_{z\to 0}\left({\rm tr} ({\rm sgn} A \vert A
\vert^{-z})\right)$. \\
The
$\zeta$-determinant of a self-adjoint operator can be expressed in terms of
the $\eta$-invariant as
follows:
\begin{prop}\label{p:etaphase} Let $A \in Ell^*_{ord >0}(M, E)$
be any self-adjoint elliptic pseudo-differential operator.
Then
\begin{equation}\label{eq:trAlogA}
{\rm tr}^A(\log A)= {\rm tr}^{\vert A \vert} (\log A)
\end{equation}
and
\begin{equation}\label{eq:detphase}
{\rm det}_\zeta (A) = \exp {\rm tr}^{\vert
A\vert}
\left(\log  A\right) ={\rm det}_\zeta
\vert A\vert\cdot  e^{{ i\pi\over
2}(\eta_A(0)-\zeta_{\vert A\vert} (0) )}.
\end{equation}
We call $\phi(A):= {\pi \over
2} \left( \eta_A(0)-\zeta_{\vert A\vert}
(0)\right) $ the phase of
${\rm det}_\zeta(A)$.
\end{prop}

{\it Proof.} Although (\ref{eq:detphase}) is a well known
result, we derive it here as a consequence of
(\ref{eq:trQ1Q2}) using the language of weighted traces.
Formula (\ref{eq:trAlogA}) relies on the fact (recalled above)
that
 res$({\rm sgn}(A))=0$.  Using  the polar
decomposition
$A=
\vert A
\vert U= U \vert A \vert$ where  $U:= {\rm sgn}(A)$  one can write
$\log A= \log \vert A \vert + \log U$ since
$[\vert A\vert, U]=0$. Applying the
results of  (\ref{eq:thmO}), we get
(with $a$ the order of $A$):
\begin{eqnarray*}
{\rm tr}^A(\log A)-{\rm tr}^{\vert A
\vert} (\log A) &=& - {a
\over 2} {\rm res}\left( (\log U)^2\right)\\
&=&
 a{\pi^2\over 8} {\rm res} ((U-I)^2)\\
&=&
a{\pi^2\over 8} {\rm res} ( U^2-2 U+I)\\
&=&
a{\pi^2\over 4} {\rm res} ( I-  U ) \\
&=&
-a{\pi^2\over 4} {\rm res} (   U )=0.
\end{eqnarray*}
In the second line we used the fact that
$U=\exp \left(  {i \pi \over 2} \left(
U-I\right) \right)$, as can easily be seen
applying either side of the equality to
eigenvectors of $A$. In the fourth line we used
the fact that $U^2=I$ since $A$ is self-adjoint,
and in the last line we used the fact that
${\rm res}(U)=0$ as proved by Atiyah, Patodi and Singer
\cite{APS I}. From this it follows that
\begin{equation}\label{eq:detodd}
{\rm det}_\zeta (A)= \exp\left(
{\rm tr}^{\vert A\vert} (\log A)\right)=
{\rm det}_\zeta  \vert A\vert e^{i\phi(A)}
\end{equation}
with $\phi(A)=-i {\rm tr}^{\vert A\vert}
\log_{({\pi
\over 2})} U= {\pi \over 2} \left(\eta_A(0)-\zeta_{\vert
A\vert}(0)\right)$. The expression in terms
of the
$\eta$-invariant follows inserting $\eta_A(0)=
{\rm tr}^{\vert A\vert} (U)$.\endsquare
 \\ \\
{\bf Remark.} This proposition yields back the definition  of
$\zeta$-determinants  for self-adjoint operators introduced by  \cite{AS},
\cite{Si}
and often used in the
physics litterature.
\\  \\
In the particular case when $A$ is (formally)
self-adjoint, the anomaly expressed in (\ref{eq:MultAn2})
vanishes:
$${\rm det}_\zeta(A^*A)={\rm det}_\zeta(A^2 )
={\rm det}_\zeta(\vert A\vert^2)=
\vert {\rm det}_\zeta(A  )\vert^2.$$
 The last equality follows from  (\ref{eq:detphase}) since
$\eta_A(0) $ and $\zeta_{\vert A \vert}(0)$ are
real.
\subsubsection*{A Pfaffian anomaly }
\begin{defin} The {\rm  Pfaffian} of
$A:= \left[ \matrix{ 0 & -D \cr
 D & 0 \cr}\right]  $ -- where
$D\in Ell^{* adm}_{ord >0} (M, E)$ is a
self-adjoint operator-- is defined by:
$${\rm Pf}_\zeta(A):= {\rm det}_\zeta(D).$$
   \end{defin}
The following result  points out to a
{\it Pfaffian anomaly} in this infinite
dimensional setting since  it shows that the
determinant is not in general the square of the
Pfaffian.
\begin{thm}\label{thm:Pfaffian} The square of the Pfaffian of
$A=
\left[
\matrix{ 0 & -D \cr
 D & 0\cr}\right]$ with $D$ self-adjoint does not
in general co\"{\i}ncide with the determinant of
$A$ for we have:
$${\rm Pf}_\zeta(A)^2= {\rm det}_\zeta(A) F_\zeta(D,
D)^{-1}={\rm det}_\zeta(A)
e^{i \pi \left(\eta_D(0)-
\zeta_{\vert D \vert} (0)\right)}$$  where
$F_\zeta(A, B)$ is the multiplicative anomaly
described in {\rm (\ref{eq:MultAn})}.
\end{thm}
{\bf Remark.} Note the fact that $e^{i \pi \left(
\eta_D(0)-
\zeta_{\vert D \vert} (0)\right)}$ is exactly the
square of the phase of the
$\zeta$-determinant of
the self-adjoint operator $D$ described in Proposition \ref{p:etaphase}. \\

{\it Proof.}
First notice that
$ \log
A  -\log \vert A \vert=- {i \pi \over 2}
\e ( iA)$,  $\e(iA):= {iA \over \vert A
\vert}$ being the sign of $iA$ where we have cut
the plane along some axis $L_\theta$ with  ${\pi
\over 2} < \theta < {3 \pi
\over 2}$.  Using this relation we can compare
$
{\rm det}_\zeta (A)$ and
${\rm det}_\zeta (\vert A
\vert)$:
\begin{eqnarray*} \log {\rm det}_\zeta (A)- \log {\rm det}_\zeta (\vert A
\vert)&=& {\rm tr}^A (\log
A)-{\rm tr}^{\vert A\vert} (\log
\vert A\vert)\cr &=& {\rm tr}^A (\log A)- {\rm tr}^{\vert A \vert} (\log
A)+ {\rm tr}^{\vert A
\vert}
\left(  \log
A  -\log \vert A \vert\right) \cr
&= &  {\pi^2 \over 8 a} {\rm res} \left( (  \e ( iA))^2 \right)
  - {i \pi \over 2}{\rm tr}^{\vert A \vert}
\left( \e ( iA)\right) \cr
 &= &  {\pi^2 \over 8 a} {\rm res} \left( I \right)
  - {i \pi \over 2}\eta_{i A} (0) \cr
 &= &   - {i \pi \over 2}\eta_{i A} (0) .
\end{eqnarray*}
Let us compute $\eta_{ iA}(0) $. If $\{\l_n , n \in \N\}$ denotes the
spectrum of
$D$, then the spectrum of $A$ is given by $\{ i\l_n, n \in \N\} \cup
\{-i\l_n, n \in \N\}$ as can be
shown considering the action of $A$ on  the orthonormal basis of
eigenvectors
$z_n:= u_n+i v_n$, $\bar z_n:= u_n-iv_n$, where $u_n:= e_n\oplus 0$, $v_n:=
0\oplus e_n$ and $e_n,
n\in  \N$ is a basis of eigenvectors of $D$ associated to the eigenvalues
$\l_n$.
Thus ${\rm tr}( A \vert A \vert^{-z-1})= i \sum_n \l_n \vert \l_n\vert^{-z-1}-i
\sum_n \l_n \vert
\l_n\vert^{-z-1}=   i{\rm tr}(D \vert D
\vert^{-z-1})+   i{\rm tr}(-D
\vert D
\vert^{-z-1})=0$ where we have used the fact that $\vert A \vert=\vert D
\vert \oplus \vert
D\vert$, and hence
$\eta_{iA}(0)=  i{\rm tr}( A \vert A \vert^{-z-1})_{\vert z=0}= 0.$ Finally we
find
$$ {\rm det}_\zeta
(A)=   {\rm det}_\zeta (\vert A \vert)
= ({\rm det}_\zeta \vert D \vert)^2.
$$
\\ We are now ready to compare ${\rm det}_\zeta (\vert A \vert)$ with
${\rm Pf}_\zeta (  A
 )^2$. Since the latter is ${\rm det}_\zeta  ( D  )^2$, it differs form
the former by the
quotient $${{\rm Pf}_\zeta (  A
 )^2\over  {\rm det}_\zeta( A)} = {\left({\rm det}_\zeta    D
\right)^2\over{\rm det}_\zeta
(\vert D
\vert)^2} =  {\left({\rm det}_\zeta    D  \right)^2\over{\rm det}_\zeta (
D^2)}= F_\zeta(D,
D)^{-1},$$ where we have used the fact that $D^2= \vert D \vert^2$ and
${\rm det}_\zeta(D^2)=
{\rm det}_\zeta ( \vert D \vert^2)=\left( {\rm det}_\zeta \vert D
\vert\right)^2$, a relation
which      can easily be derived from the triviality of the multiplicative
anomaly $F_\zeta(\vert
D\vert,
\vert D \vert)$.
 \endsquare
\section{Variations of $\eta$-invariants as
integrated trace anomalies}
Given  two invertible self-adjoint elliptic
operators $A_1$ and $A_0$, the spectral flow of a family of self-adjoint
elliptic operators  $\{A_t, t\in [0, 1]\}$
interpolating them measures the net number of
times the spectrum $\bigcup_{t\in [0, 1]}
{\rm Spec} (A_t)$ of the family $\{A_t, t\in [0, 1]\}$ crosses the zero
axis   \cite{APS III}.
  Making this definition precise requires
some care since there might well be an infinite
 number of crossings of the zero
axis. There are different ways of defining the
spectral flow see e.g.
\cite{BLP}, \cite{Me}. Let us
first observe that \cite{Me}
\begin{lem}\label{l:melrose}
There is  a partition
$t_0=0<t_1<\cdots<  t_N =1$ of the interval
$[0, 1]$ and there are real numbers $\l_i, i= 1,
\cdots, N$, $\l_0=\l_{N+1}=0$ such that the
spectrum of
$ A_t $  avoids $\l_i$ for any $t$ in the interval
$[t_i, t_{i+1}]$.
\end{lem}

{\it Proof.} It follows from the
discreteness of the spectrum ${\rm Spec}(A_t)$ of
$A_t$ that, given any
$t_0\in ]0, 1[$,  there is some
$\l_0\in\R$ which avoids the spectrum of
$A_{t_0}$. For $\l\in \R$, let  $U_{\l }:= \{t\in
]0, 1[,\l \notin {\rm Spec}(A_t)\} $. From the
the continuity of the family $\{A_t\}$, it
follows that
$U_\l$ is an open subset of $]0, 1[$.
Since $U_{\l_0}$   contains
$t_0$, it also contains  the closure of some open
interval
$I_{\l_0}$ centered at  $t_0$. It is clear from the
construction  that
$[0, 1]
\subset
\bigcup_{\l\in
\R} \bar  I_\l$.    Since
$[0, 1]$ is  compact,  one can extract from this
covering a finite covering
$I_{\l_i}:= [t_{i-1}, t_{i }], i=1, \cdots, N $
where $\l_0=\l_{N+1}=0$ (recall that $A_0$ and
$A_1$ are invertible),
$t_0=0<t_1<
\cdots<  t_{N }=1$, such that $\l_i$ does not
belong to  $\{{\rm Spec}(A_t), t\in [t_{i-1}, t_{i
}]\}$.   \endsquare
\\ \\
Let $t_i, i=0, \cdots, N $, $\l_j, j=0,\cdots,
N+1$ be as in the above lemma. The spectral flow
of the family $\{A_t\}$  is defined by
\cite{Me} (formula (8.134)):
\begin{equation}
 {\rm SF} (\{A_t\})  :=
 \sum_{i=0}^N \sum_{\l\in
{\rm Spec}(A_{t_i})\cap [\l_i, \l_{i+1}]
}\hbox{sgn}(\l_{i+1}- \l_i) m(\l, t_i) , 
\end{equation}
\\
where $m(\l, t)$ denotes the multiplicity of $\l$
in the spectrum of $A_t$ and ${\rm sgn}(\alpha)$ is
$-1, 0$ or $1$ as $\alpha$ is negative, $0$ or
positive. One can check that this definition  is
independent of the chosen partition. It also follows
from the definition that if $A_t$ is invertible
for any $t\in [0, 1]$, then ${\rm SF} (A_t)=0$ as expected.
As a further consequence of the definition,
given
$\alpha \in \R$, then
\begin{equation}\label{eq:sfshift}
 {\rm SF} (\{A_t -\alpha\})  := {\rm SF}
(\{A_t  \})  + {\rm sgn} (\alpha ) \left[{\rm tr} (P_{1,
\alpha}) -{\rm tr} (P_{0,
\alpha}) \right] .
\end{equation}
(Compare with
formula (8.135) in
\cite{Me}). Here $P_{t, \alpha}$ denotes the
orthogonal projection onto the finite dimensional
space generated by eigenvectors of $A_t$ with
eigenvalues in $[0, \alpha]$ or $[\alpha, 0]$, according to whether
$\alpha$ is positive or negative.
\\
\\ In order to relate the difference of the
$\eta$-invariants $\eta_{A_1} (0)-
\eta_{A_0}(0)$ to the  spectral flow, we need the
 following
\begin{lem}\label{l:trsign0}
Let
$\{A_x, x\in X\}$ be a  smooth family of
self-adjoint elliptic operators with constant
positive order
$a$ parametrized by some manifold $X$. On an open
subset   of $X$  where
  the map $x \mapsto A_x$ is invertible, the map  $ x
\mapsto \tr^{\vert A_x\vert }({\rm sgn} ( A_x))$,
where ${\rm sgn} (A_x):= A_x
\vert A_x \vert^{-1}$,
is differentiable and   we
have:
\begin{equation}\label{eq:trsign0}
 d\,  \left( {\rm tr}^{\vert A \vert }
 \left( {\rm sgn} (A )\right)\right)
= [d\, , {\rm tr}^{\vert A\vert}] \left( {\rm sgn}
(A )\right)=- {1 \over
a}{\rm res}      (A^{-1} {d  }
 \vert A  \vert )= - {1 \over
a}{\rm res}      (\vert A\vert ^{-1} d A )
\end{equation}
where we have set $ [d,{{\rm tr}}^{Q} ]:= d\circ
{\rm tr}^{Q}-{\rm tr}^{Q} \circ d$.
\end{lem}

{\it Proof.} On one hand it follows from
(\ref{eq:dtr}) that
\begin{eqnarray*}
  \left[d\, , {\rm tr}^{\vert A \vert }
\right] ({\rm sgn} (A ) )
&= &  - {1 \over a}{\rm res} ({\rm sgn} (A )
{d }\log
\vert A
\vert)\\
&=& - {1 \over a}{\rm res} ({\rm sgn} (A ) \vert A
\vert^{-1} {d  } \vert A
\vert )\\
&= & - {1 \over a}{\rm res}      (A^{-1} {d  }
 \vert A  \vert )
\end{eqnarray*}
where we have used the fact that $[\vert A \vert,
{\rm sgn} A ]=0$. On the other hand, by \cite{APS
III}, Proposition (2.10), we have:
\begin{eqnarray*}
 {d } \left( {\rm tr}^{\vert A\vert  } ({\rm sgn}
(A ) )
\right)& =& {d  } \left(
{\rm f.p.} \left( {\rm tr} ({\rm sgn} A  \vert A
\vert^{-z}\right) \right)_{\vert_{z=0}}\\
&=& -{\rm f.p.} \left(z \left(
 {\rm tr} (d A \vert A  \vert^{-(z+1)}\right)
\right)_{\vert_{z=0}}\\ &=& -{1 \over a}{\rm res}
\left( d A  \vert A  \vert^{-1} \right) .
\end{eqnarray*}
But by proposition (2.11) in \cite{APS III}, the
map   res(sgn$(A )$) is constant for a continuous
variation of $A$ and hence
 \begin{eqnarray*}
{\rm res} (A^{-1} d \vert A\vert )
- {\rm res} (dA \vert A \vert^{-1})
 &=&
d\,{\rm res} (A^{-1} \vert A \vert) \\
&=&  d\, {\rm res} ({\rm sgn}A )\\
&=&  0
\end{eqnarray*}
so that finally
$$
   d\,  \left( {\rm tr}^{\vert A \vert }
 \left( {\rm sgn} (A )\right)\right)
= [d\, , {\rm tr}^{\vert A\vert}] \left( {\rm sgn}
(A )\right)=- {1 \over
a}{\rm res}      (A^{-1} {d  }
 \vert A  \vert )= - {1 \over
a}{\rm res}      (\vert A\vert ^{-1} d A )
$$
as claimed in the lemma.
\endsquare
\\ \\
The following theorem relates the
variation of $\eta$ invariants to an
integrated trace anomaly. \\
\begin{thm}\label{th:etatrace} Let
$\{A_t, t\in [0, 1]\}$ be a smooth family of
self-adjoint  invertible  elliptic
operators  with constant order in $C l(M, E)$.
Then
 \begin{eqnarray*}
\eta_{A_1}(0)- \eta_{A_0}(0)&= &   \int_{0}^{1}
\dot
{{\rm tr}}^{\vert A_t\vert } ({\rm  sgn} (A_t))dt
\nonumber
\\ &= &  - {1\over a}
 \int_0^1 {\rm res} \left( \dot A_t
\vert A_t
\vert^{-1}\right) dt
\end{eqnarray*}
which
relates the difference of the $\eta$-invariants
$ \eta_{A_1}(0)- \eta_{A_0}(0)$ to  an integrated
trace anomaly $\int_{0}^{1}
\dot
{{\rm tr}}^{\vert A_t\vert } ({\rm  sgn}
(A_t))dt$  where  we have set $\dot {{\rm
tr}}^{\vert A_t\vert }:= {d
\over dt}{{\rm tr}}^{\vert A_t\vert }$.
\end{thm}

{\it Proof.} Applying the first
identity in  (\ref{eq:trsign0}) to a family parameterized by $[0,
1]$ yields
$${d\over dt} \eta_{A_t}(0)=  {d\over dt} \left(
\hbox{tr}^{\vert A_t\vert}
(\hbox{sgn}(A_t))\right)=  {\dot
{\rm tr}}^{\vert A_t\vert} \left({\rm sgn}
(A_t)\right)$$ and hence
$$\eta_{A_1}(0)-\eta_{A_0}(0)= \int_0^1 {d \over
dt} \eta_{A_t}(0)dt=
 \int_0^1 {\dot
{\rm tr}}^{\vert A_t\vert} \left({\rm sgn}
(A_t)\right)dt.$$
\endsquare
 \\ \\
The following corollary of Theorem
\ref{th:etatrace} is a reformulation of the
Atiyah-Patodi-Singer theorem in terms of weighted
trace anomalies. We derive it from Theorem
\ref{th:etatrace}, closely following the proof of
Proposition 8.43 in 
\cite{Me}.
\begin{cor}\label{cor:etatrace} Let
$A_1$ and
$A_0$ be two invertible elliptic self-adjoint
  operators with common order $a $   and
let
$\{A_t, t\in [0, 1]\}$ be a smooth family of
self-adjoint (possibly non invertible) elliptic
operators  with fixed order $a$ 
 interpolating
$A_0$ and
$A_1$. Then
\begin{eqnarray}\label{eq:etatrace}
\eta_{A_1}(0)- \eta_{A_0}(0) & = & 2 {\rm  SF}
(\{A_t\})+ \int_{0}^{1}
{\dot{\rm tr}}^{\vert A_t\vert} ({\rm  sgn} (A_t))dt \nonumber
\\ & = & 2 {\rm  SF} (\{A_t\})- {1\over a}
 \int_0^1 {\rm res} \left( {\dot A_t}
\vert A_t
\vert^{-1}\right) dt ,
\end{eqnarray}
which
relates the difference of the $\eta$-invariants
$ \eta_{A_1}(0)- \eta_{A_0}(0)$ to
 the spectral flow, via an integrated trace anomaly
 $ \int_{0}^{1}
\dot
{{\rm tr}}^{\vert A_t\vert } ({\rm  sgn} (A_t))dt$
involving $\dot {{\rm tr}}^{\vert A_t\vert }({\rm
sgn} A_t)$.
  \end{cor}
{\bf Remark.} The residue on the r.h.s. of
(\ref{eq:etatrace}) corresponds to the local
term
$\int_0^1 \dot
\eta^c_{A_t}(0) dt $ --where $\eta^c_{A_t}$ is the
``continuous" part of the $\eta$-invariant-- which
 arises in the
Atiyah-Patodi-Singer theorem for a family of
self-adjoint Dirac operators
$\{A_t\}$. In other words we have the following
schematic correspondence: \\
$$
\begin{array}{ccccc}
{\begin{tabular}{|p{5.2cm}|}
\hline local term in the
Atiyah-Patodi-\\Singer theorem for families\\
\hline
\end{tabular}} & \Leftrightarrow &
{\begin{tabular}{|p{2.7cm}|}
\hline an
integrated \\ {\it tracial}
anomaly\\
\hline
\end{tabular}}
\end{array}
$$
\\

{\it Proof of the Corollary.} We  show that
one can reduce the proof of the Corollary to the
case of a family of invertible operators, and then
apply Theorem
\ref{th:etatrace} which yields the
desired formula in that case.   In  order to
reduce the proof to the case of a family of
invertible operators, let  us first observe that
formula (\ref{eq:etatrace}) is invariant under a
shift
$A_t \mapsto A_t- \alpha$,
$\alpha \in
\R$. Let us first consider the case $\alpha\geq
0$.
Since
$A_t$ has only a finite number of eigenvalues
(counted with multiplicity) contained in
$[0, \alpha]$, under   a
shift $A_t\mapsto A_t- \alpha$ its $\eta$ invariant
will change  by minus this number of eigenvalues
and  we have
$$
 \eta_{A_t-\alpha}(0)=
\eta_{A_t}(0)-\tr(P_{t, \alpha}),
$$
where as before $P_{t,
\alpha}$ denotes the orthogonal projection onto
the finite dimensional space generated by
eigenvectors of $A_t$ with eigenvalues in $[0,
\alpha]$ or $[\alpha, 0]$ according to the sign
of $\alpha$.
In a similar way, for $\alpha\leq 0$ we have
$$
 \eta_{A_t-\alpha}(0)=
\eta_{A_t}(0)+\tr(P_{t, \alpha}),
$$
and hence for any $\alpha \in \R$
\begin{equation}\label{eq:etashift}
 \eta_{A_t-\alpha}(0)=
\eta_{A_t}(0)+ {\rm sgn}(\alpha)\tr(P_{t, \alpha}).
\end{equation}
As a consequence, we find:
\begin{equation}\label{eq:etashifts}
\eta_{A_1-\alpha}(0)-
\eta_{A_0-\alpha}(0)=
\eta_{A_1}(0)- \eta_{A_0}(0)+
 {\rm sgn}(\alpha) \left[ \tr(P_{1, \alpha}) - \tr(P_{0, \alpha}) \right].
\end{equation}
Let us now investigate how ${\rm res} (\dot A_t
\vert A_t\vert^{-1})$ changes under such a
shift.
From  Lemma \ref{l:trsign0} it follows that
for $\alpha\in
\R$ then :
\begin{eqnarray*}
 \dot {\rm tr}^{ \vert A_t-\alpha\vert}
 \left({\rm sgn}( A_t-\alpha)\right)-
\dot {\rm tr}^{ \vert A_t \vert}
\left({\rm sgn}( A_t )\right)&= & { d \over dt}\,
\left( \eta_{A_t-\alpha }(0) -
\eta_{A_t}(0)\right)\nonumber \\
   &=& -{\rm sgn}(\alpha){d \over dt} \hbox{tr}
(P_{t, \alpha}),
\end{eqnarray*}
and hence that:
\begin{equation}\label{eq:resshift}
\int_0^1 \left[ \dot {\rm tr}^{ \vert A_t-\alpha\vert}
 \left({\rm sgn}( A_t-\alpha)\right)-
\dot {\rm tr}^{ \vert A_t \vert}
\left({\rm sgn}( A_t )\right) \right] \, dt =
-\hbox{sgn}(\alpha)\left[\tr(P_{1, \alpha})-
\tr(P_{0,\alpha})\right].
\end{equation}
Combining  formulae
(\ref{eq:etashift}), (\ref{eq:resshift}) and
(\ref{eq:sfshift}), giving the  variation of the
various ingredients of formula (\ref{eq:etatrace})
 under a shift by
$\alpha$, shows that  a shift of the family of
operators   by
$\alpha$ does not modify equation
(\ref{eq:etatrace}).
\\
Using the
partition of $[0, 1]$ introduced in Lemma
\ref{l:trsign0},     equation  (\ref{eq:etatrace})
can be seen as a combination of the following
equations:
$$\eta_{A_{t_i}}(0)- \eta_{A_{t_{i-1}}}(0)=
2 {\rm SF} \left(
\{A_t\}_{t\in [t_{i-1}, t_i]}\right)+
\int_{t_{i-1}}^{t_i}{\dot {\rm tr}}^{\vert
A_t\vert} ({\rm sgn}(A_t))dt, \;\;\;\; i=1, \cdots,
N.  $$
By the above preliminary remarks, it
suffices to show this for any shift
$A_{t_i}-\alpha $ of
$A_{t_i}$.  Since by Lemma \ref{l:melrose}  we
know the existence of
$\l_i\in
\R, i=1,
\cdots, N$ such that ${\rm Spec}(A_t-\l_i)$ does
not meet the zero axis on $[t_{i-1}, t_i]$, the 
proof of the theorem indeed reduces to
the case when all the  operators in the family
$\{A_t\}$  are invertible, considered in Theorem \ref{th:etatrace}.
\endsquare
\\ \\
As we shall see later on,
 the Atiyah-Patodi-Singer index theorem gives an explicit description of
the local term arising
from the Wodzicki residue in (\ref{eq:etatrace}) for classes of
Dirac operators. As a consequence of the above discussion we have:
\begin{cor}\label{c:phaseres} Let $M$ be an odd
dimensional manifold and let
$\{A_t, t\in [0, 1]\}$ be a smooth family of
 self-adjoint elliptic
pseudo-differential operators of positive constant order $a$ with  vanishing
spectral flow interpolating two   invertible differential (more generally
odd-class
pseudo-differential) operators
$A_0, A_1$. Then  the difference of phases
$\phi(A_1)- \phi(A_0)$  of the $\zeta$-determinants of $A_1$ and $A_0$ can
be expressed in terms of
an integrated weighted trace anomaly involving
$\dot {\rm tr}^{\vert A_t \vert} ({\rm sgn} (A_t))$:
\begin{eqnarray}\label{eq:phaseres}
\phi(A_1)- \phi(A_0) &= & {\pi \over 2} \left( \eta_{A_1}(0)-
\eta_{A_0}(0)\right) \nonumber \\
&=&  {\pi \over 2}
\int_{0}^{1} {\dot  {\rm  tr}}^{\vert
A_t\vert} ({\rm sgn} (A_t)) dt \\
&=& - {\pi \over 2a}   \int_{0}^{1} {\rm
res} (\vert A_t
\vert^{-1} \dot A_t )
dt. \nonumber
\end{eqnarray}
 where we have kept the notations of
 Theorem {\rm \ref{th:etatrace}}.
\end{cor}

{\it Proof.}  The phase $\phi(A_i)$, $i=0,1$ given by (see (\ref{eq:detodd})) 
$\phi(A_i)= {\pi \over 2} \left( \eta_{A_i} (0) 
- \zeta_{\vert {A_i}\vert} (0) \right)$ 
reduce here to ${\pi \over 2}  \eta_{A_i} (0) $ since 
$ \zeta_{\vert {A_i}\vert} (0)= 0$ vanishes in odd dimensions \cite{Si}. 
\endsquare
\section{Determinant bundles
and trace anomalies}
We first need to recall the construction of
determinant bundles for families of elliptic
operators on closed manifolds.  We shall not
recall it in full detail, referring the reader
to \cite{Q1}, \cite{BF} for a precise description  of the
local trivializations involved in the
construction of the determinant bundle. In
order  to avoid technicalities, here we only
state the results at points for which the
operator is invertible, which simplifies
the presentation of the formulae. Let
$\M\to X$ be a smooth (locally trivial) fibration
of manifolds based on a smooth manifold
$X$ modelled on some closed Riemannian manifold
$M$. Let
$ \E^+\to \M$, resp. $\E^-\to
\M$,
  be a Hermitian finite rank  vector
bundle   on
$\M$ and let
$  {\cal E}^+ \to X$, resp.  ${\cal E}^-\to X$   be
the induced infinite rank superbundle on $X$ with
fibre above
$x$ given by ${{\cal E}}_x^+:= \Ci(M_x, E_x^+) $,
resp.
${{\cal E}}_x^-:= \Ci(M_x, E_x^-) $,
$M_x,$, resp. $E_x^+ $, resp. $E_x^- $ being the
fibre of
$\M$, resp. of
$\E^+$, resp. $\E^-$ above
$x$.
\subsubsection*{A metric on ${\cal E}^+$, ${\cal
E}^-$ }
The Hermitian metric  on
$\E^+$, resp.
$\E^-$   induces   a metric on ${\cal E}^+$, resp.
${\cal E}^-$  :
\begin{equation}\label{eq:herm+}
\langle \sigma^+, \rho^+ \rangle_x^{{\cal E}^+}
=
\int_{\M
/ X} \langle \sigma^+ (x), \rho^+(x)
\rangle_m^{E_x^+} d\mu_x(m),
\end{equation}
resp.
\begin{equation}\label{eq:herm-}
\langle \sigma^-, \rho^- \rangle_x^{{\cal
E}^-} =
\int_{\M
/ X} \langle \sigma^- (x), \rho^-(x)
\rangle_m^{E_x^-} d\mu_x(m)
\end{equation}
 where $\mu_x(m)$ is the volume element on
the fibre $M_x$, $\sigma^+, \rho^+ \in \Ci(X,
{\cal E}^+)$, resp. $\sigma^-, \rho^- \in \Ci(X,
{\cal E}^-)$ and $\langle  \cdot, \cdot
\rangle^+_m$, resp. $\langle  \cdot, \cdot
\rangle^-_m$ are Hermitian products on the fibres
  $E^+_m$ and $E^-_m$.
\subsubsection*{A connection on ${\cal E}^+$,
${\cal E}^-$  }
Given a horizontal distribution on
$\E^+$, resp. $\E^-$, one can build a connection
$\tilde \nabla^{{\cal E}^+}$, resp.
$\tilde \nabla^{{\cal E}^-}$ on
${\cal E}^+, {\cal E}^-$ from a connection
$\nabla^{\E^+}$, resp. $\nabla^{\E^-}$ on
$\E^+$, resp.  $\E^-$:
\begin{equation}\label{eq:conn+}
\left(\tilde \nabla_U^{{\cal E}^+}\sigma
\right)(m):=
  \nabla_{\tilde U(m)}^+
\sigma(x),
\end{equation}
resp.
\begin{equation}\label{eq:conn-}
\left(\tilde \nabla_U^{{\cal E}^-}\sigma
\right)(m):=
\nabla_{\tilde U(m)}^- \sigma(x)
\end{equation}
 where $U\in T_xX$ and $\tilde U(m)$ is the
horizontal lift of $U$ at point $m\in M_x$. \\
This connection   needs to be slightly
modified to become compatible with the above
metric on ${\cal E}$ \cite{BGV}, \cite{BF}:
$$\nabla^{{\cal E}^+} :=\tilde \nabla^{{\cal
E}^+}   + {1 \over 2} {\rm div}_{M_x}
 $$
resp. $$\nabla^{{\cal E}^-} :=\tilde
\nabla^{{\cal E}^-} + {1 \over 2}
{\rm div}_{M_x}
$$
 where ${\rm div}_{M_x}$ is the divergence of
the volume form in the direction of the base
manifold
$X$.
\subsubsection*{The Quillen determinant
bundle }
Let
$\{A_x^+:{\cal E}^+_x\to {\cal E}^-_x, x\in X\}$ be
a smooth family of
  elliptic admissible operators with
constant positive order $a$. They yield a smooth
family of Fredholm operators
$\{ A_x^{+ ,s}: H^s (M_x, E_x^+) \to
H^{s-a} (M_x, E_x^-), x\in X\}$ with $s\in \R$.
Following Quillen \cite{Q1}, to this family of Fredholm
operators, we can associate a determinant bundle
${\cal L}_{A^+}$.
\\  There is a metric on ${\cal L}_{A^+}$ called
the Quillen metric \cite{Q1} defined at a point $x$
where
$A_x^+$ is invertible by:
\begin{equation}\label{eq:Qmetric}
\Vert Det A^+ \Vert_x^Q:=  {\rm det}_\zeta
 \vert A_x^+\vert
\end{equation}
where $Det A^+$
is a section of
${\cal L}_{A^+}$.
\subsubsection*{A connection on the
determinant bundle}
 Following \cite{BF} let us now
 equip the
determinant bundle with a connection. It arises as
a  natural extension of the well-known formula for
the logarithmic variation of the determinant of  a
family of invertible elliptic operators, which we
recall here and prove using the language of
weighted traces.
\begin{lem}\label{l:dlogdet} Let
$E\to M$ be a fixed Hermitian vector bundle over a
fixed closed Riemannian manifold.  Let
$A_x\in Ell^{adm}_{ord>0} (M, E)$ be a smooth
family parametrized by some smooth manifold $X$
with a common spectral cut and constant order $a$.
Then, at a point $x\in X$ at which $A_x$ is
invertible we have for $h\in T_xX$:
\begin{equation}\label{eq:dlogdet}
d \log
{\rm det}_\zeta( A  )(h)=
{\rm tr}^{A_x}( A_x^{-1} d A(h) ).
\end{equation}
\end{lem}

 {\it Proof.} Let $\{\gamma_x(t), t\in [0,
t_0]\}$ be a curve on $X$ driven  by $h$ and
starting at
$x$ at time $t=0$.
\begin{eqnarray*} d
\log
{\rm det}_\zeta( A )(h)&=& d {\rm tr}^{A } (
\log A )(h)\\
&=& {\rm tr}^{A_x} (d \log
A )(h)+ [d, {\rm tr}^{A }](h) (\log A_x)\\
 &=& {\rm tr}^{A_x} (A_x^{-1}d A(h)) + \lim_{
t\to 0} t^{-1}
\left({\rm tr}^{A_{\gamma_x(t)}}(\log A_x)-
{\rm tr}^{A_x} (\log A_x)\right)\\
&=& {\rm tr}^{A_x} (A_x^{-1}d A(h)) -{1 \over 2}
\lim_{ t\to 0}  t^{-1}
{\rm res}\left(  (\log A_{\gamma_x(t)} -
 \log A_x)^2 \right)\\
&=& {\rm tr}^{A_x} (A_x^{-1}d A(h))
\end{eqnarray*}
 where we have used formula (\ref{eq:thmO}). \endsquare
   \\ \\
$\bullet$ When $\E^+= \E^-= \E$,
setting
${\cal E}:= {\cal E}^+= {\cal E}^-$ and letting
$\{A_x:= A_x^+, x\in X\}$ be a family of formally
self-adjoint operators, the above computation gives
a hint for the choice of  a
 connection on
${\cal L}_A$. We define it    at a point
$x\in X$ where
$A_x$ is invertible by:
\begin{equation}\label{eq:Nablalog}
\left(Det A_x \right)^{-1}
\nabla^{Det} Det A := {\rm tr}^{A_x }
\left(  A_x^{-1}
[\nabla^{\cal E} , A ] \right).
\end{equation}
\\ This connection is compatible with the
Quillen metric as the following lemma shows:
\begin{lem}\label{Retr1} Let $\{A_x, x\in X\}$ be a family
of formally self-adjoint elliptic operators and
${\cal L}_A$ the associated determinant bundle on
$X$. The connection {\rm (\ref{eq:Nablalog})} is compatible with the
Quillen metric. Namely:
 $${\cal R} e \left({\rm tr}^{  A_x   }
\left( A_x^{-1}
\nabla^{Hom({\cal E})}A \right)  \right)= d \log
\Vert Det A
\Vert_Q  $$
at a point
$x$ where
$A_x$ is invertible.\\
Moreover the imaginary part coincides with an
infinitesimal tracial anomaly:
$${\cal I}m \left({\rm tr}^{  A_x   }
\left( A_x^{-1}
\nabla^{Hom({\cal E})}A \right)  \right)={\pi
\over 2} [\nabla, \tr^{\vert A\vert}]({\rm sgn}
A -I)= -{\pi \over 2} {\rm res} \left( ({\rm sgn
} A-I)\vert A\vert^{-1} [\nabla, \vert
A\vert]\right),
$$
a tracial anomaly combining {\rm (\ref{eq:coboundary})} and 
{\rm (\ref{eq:dtr})}.
\end{lem}

{\it Proof.} Writing
$\nabla^{\cal E}= d + \theta^{\cal
E}$ locally, it follows from (\ref{eq:dlogdet})
that:
\begin{eqnarray*} d\log {\rm det}_\zeta(A)&=&
{\rm tr}^A(A^{-1} dA)\\
&=&  {\rm tr}^A(A^{-1} dA)+ {\rm tr}^A(A^{-1}
[\theta^{\cal E}, A])\\
&=&   {\rm tr}^A(A^{-1}
[\nabla^{\cal E}, A])= {\rm tr}^A(A^{-1}
 \nabla^{ Hom {\cal E}}( A )).
\end{eqnarray*}
Thus,
differentiating (\ref{eq:detphase}) yields:
 $$ {\rm tr}^{  A_x   }
\left( A_x^{-1}
\nabla^{Hom({\cal E})}A \right)= d\log
{\rm det}_\zeta \vert A\vert+ {i \pi \over 2}
d\, \left(\eta_A(0)- \zeta_{\vert A
\vert}(0)\right). $$ 
Since $\eta_A(0)- \zeta_{\vert
A \vert}(0)$  is real, the first part follows using
(\ref{eq:Qmetric}) with $A^+=A$. \\
As for the second part of the lemma, we
have:
 \begin{eqnarray*}{\cal I}m \left({\rm tr}^{
A_x   }
\left( A_x^{-1}
\nabla^{Hom({\cal E})}A \right)\right)&=& {\pi
\over 2} d {\rm tr}^{ \vert  A_x  \vert }
\left({\rm sgn} A-I \right) \\
&=&  {\pi \over
2}\left[ d, {\rm tr}^{ \vert  A_x  \vert }\right]
\left({\rm sgn} A-I \right) \quad \hbox{ by
formula (18)}\\ &=&  {\pi \over
2}\left[ \nabla^{{\cal E}}, {\rm tr}^{ \vert  A_x
\vert }\right]
\left({\rm sgn} A-I \right)
\end{eqnarray*}
 where we have used the fact that  $$ {\rm tr}^{
\vert  A_x
\vert }\left([\theta^{\cal E}, {\rm sgn} A-I
]\right)= -{1
\over a} {\rm res} ([\log \vert A\vert ,
\theta^{\cal E}] ({\rm sgn} A-I )])= -{1 \over a}
{\rm res} ([{\rm sgn}
A-I  , \log \vert A\vert ] \theta^{\cal E} )=
0.$$  Here as before,
$\theta^{\cal E}$ is the local one form arising in
a local description of the connection
$\nabla^{\cal E} $.
\endsquare
 \\ \\
$\bullet$  When $\E^+\neq  \E^- $,
letting $\E:= \E^+\oplus \E^-$ be the
finite rank supervector bundle built from the
direct sum, and
${\cal E}:= {\cal E}^+\oplus {\cal E}^-$ the
corresponding infinite rank supervector bundle,
following
  Bismut and Freed  \cite{BF}, we equip  the bundle
${\cal L}_{A^+}$ with a connection whose
expression is a generalization of the r.h.s.
of (\ref{eq:dlogdet}) up to the fact that the weight $A_x$ is
replaced by $\vert A_x\vert$. At a point
$x$ at which
$A_x^+$ is invertible, the Bismut-Freed connection
reads:
\begin{equation}\label{eq:sNablalog}
\left(Det A_x^+\right)^{-1}
\nabla^{Det} Det A^+:= {\rm tr}^{\vert A^+_x
\vert}
\left(\left(  A_x^+\right)^{-1}
\nabla^{Hom({\cal E}^+, {\cal E}^-)}
A^+ \right).
\end{equation}
\begin{lem}\label{l:Retr2} {\bf \cite{BF}}  Let $\{A_x^+, x\in
X\}$ be a family of   elliptic
operators and
${\cal L}_{A^+}$ the associated determinant bundle
on
$X$. The Bismut-Freed
connection  is    compatible with the Quillen
metric, namely
 $${\cal R} e \left({\rm tr}^{\vert A_x^+ \vert}
\left(\left(  A_x^+\right)^{-1}
\nabla^{Hom({\cal E}^+, {\cal
E}^-)}A^+ \right)\right)= d \log \Vert Det A^+
\Vert_Q  $$ at a point $x$ where
$A_x^+$ is invertible.
\end{lem}

 {\it Proof.}
\begin{eqnarray*} &2&{\cal R} e \left({\rm tr}^{\vert
A_x^+
\vert}
\left(\left(  A_x^+\right)^{-1}
\nabla^{Hom({\cal E}^+, {\cal E}^-)}
A^+ \right)\right)\\
 &=&
 {\rm tr}^{\vert A_x^+ \vert}
\left(\left(  A_x^+\right)^{-1}
\nabla^{Hom({\cal E}^+, {\cal E}^-)}A^+ \right) +
\left({\rm tr}^{\vert A_x^+
\vert}
\left(\left(  A_x^+\right)^{-1}
\nabla^{Hom({\cal E}^+, {\cal
E}^-)}A^+ \right)^*\right) \\
&=&{\rm tr}^{\vert A_x^+ \vert}
\left(\left(  A_x^+\right)^{-1}
\nabla^{Hom({\cal E}^+, {\cal E}^-)}A^+ \right) +
 {\rm tr}^{\vert A_x^+
\vert}
\left(  \nabla^{Hom({\cal E}^+, {\cal E}^-)}
A^- \left(  A_x^-\right)^{-1}\right)  \\
&=& {\rm tr}^{\vert A_x^+ \vert}
\left(\left(  A_x^+\right)^{-1}
\nabla^{Hom({\cal E}^+, {\cal E}^-)}A^+ \right) +
\left({\rm tr}^{\vert A_x^-
\vert}
\left( \left(
A_x^-\right)^{-1} \nabla^{Hom({\cal E}^+, {\cal
E}^-)} A^- \right) \right)\\
&=& {\rm tr}^{\vert A_x^+ \vert}
\left(\left(A_x^- A_x^+\right)^{-1}
\nabla^{Hom({\cal E}^+ )}
A^-A^+ \right)  \\
&=&   2{\rm tr}^{\vert A_x^+
\vert}
\left( \vert   A_x^+\vert^{-1}
 \nabla^{Hom({\cal E}^+ )} \vert
A^+\vert \right)  \\
&=& 2{\rm tr}^{\vert A_x^+ \vert}
\left( \vert   A_x^+\vert^{-1}
d\vert A^+ \right)
+ 2{\rm tr}^{\vert A_x^+ \vert}
\left( \vert   A_x^+\vert^{-1}
[\theta^{{\cal E}^+}, \vert A^+\vert]\right)  \\
&=& 2{\rm tr}^{\vert A_x^+ \vert}
\left( \vert   A_x^+\vert^{-1}
d\vert A^+ \right)  \\
&=& 2d \log {\rm det}_\zeta  \vert A^+ \vert
\end{eqnarray*}
where we have set $A^-:= \left(A^+\right)^*$ and
written $\nabla^{{\cal E}^+}= d + \theta^{{\cal
E}^+} $ locally.
\endsquare
\\ \\
 Note that one could also have equipped the
bundle ${\cal L}_A$ with the Bismut-Freed
connection in the self-adjoint case,  which would
amount to taking the weight $\vert A\vert$
instead of the weight $A$ chosen in formula
(\ref{eq:Nablalog}).
\begin{lem}\label{l:BFcreal} In the self-adjoint case, the Bismut-Freed
connection
$$ \left(Det A_x \right)^{-1}
\tilde \nabla^{Det} Det A:={\rm tr}^{\vert A_x
\vert}
\left( A_x^{-1}
[\nabla^{ {\cal E}} ,
A]\right)  = d\log
{\rm det}_\zeta(\vert A \vert)$$ is a purely
real exact form given by the exterior differential
of the Quillen metric.
\end{lem}

 {\it Proof.}
 The result follows from the
fact that
$\overline{ {\rm tr}^{\vert A \vert} (B)}=
{\rm tr}^{\vert A \vert} (B^*)$ as  the
following computation shows:
\begin{eqnarray*}2{\cal I}m \left( \left(Det A_x
\right)^{-1}
\tilde \nabla^{Det} Det
A\right)& =&  {\rm tr}^{\vert A_x
\vert}
\left( A_x^{-1}
\nabla^{Hom({\cal E})}
A \right) -\overline{  {\rm tr}^{\vert A_x
\vert}
\left( A_x^{-1}
\nabla^{Hom({\cal E})}
A \right)}\\
&=&   {\rm tr}^{\vert A_x
\vert}
\left( A_x^{-1}
\nabla^{Hom({\cal E})}
A \right)- {\rm tr}^{\vert A_x \vert}
\left(
\nabla^{Hom({\cal E})}
A^* \left(A_x^{-1} \right)^*\right) \\
&=&   {\rm tr}^{\vert A_x
\vert}
\left( A_x^{-1}
\nabla^{Hom({\cal E})}
A \right)- {\rm tr}^{\vert A_x
\vert}
\left(
 A_x^{-1}
\nabla^{Hom({\cal E})}
A \right) \\
&=&0.
\end{eqnarray*}
\endsquare
\subsubsection*{The curvature on the
determinant bundle }
The following theorem
relates the curvature on the determinant bundle
  to trace anomalies.
\begin{thm}
\begin{enumerate}
\item  When $\E^+= \E^-= \E$, setting
${\cal E}:= {\cal E}^+= {\cal E}^-$ and letting
$\{A_x:= A_x^+, x\in X\}$ be a family of formally
self-adjoint operators, the connection  differs
from an exact form by a trace anomaly  of type
{\rm (\ref{eq:trQ1Q2})}:
\begin{eqnarray}\label{eq:Nablares} \left( Det A_x \right)^{-1}
\nabla^{Det} Det A &=&  d\log
{\rm det}_\zeta(\vert A \vert)+
\left({\rm  tr}^{  A_x }- {\rm tr}^{\vert
A_x\vert} \right)\left(A_x^{-1} [\nabla^{\cal E},
A] \right)  \\  &=&d\log
{\rm det}_\zeta(\vert A \vert) -{1 \over a}
 {\rm res}
\left( A_x^{-1} [\nabla^{{\cal E}}  , A] (\log
A_x-
\log \vert A_x \vert)\right). \nonumber
\end{eqnarray}
 In particular, the curvature is a
differential of a trace anomaly residue.
\item  When
$\E^+\neq
\E^-
$, letting
$\E:= \E^+ \oplus \E^-$ and  ${\cal E}:= {\cal
E}^+\oplus {\cal E}^-$,  we set
$ A :=
\left[
\matrix {0& A^+ \cr
A^- & 0 \cr } \right]$ with $A^-$ the formal
adjoint of $A^+$.
  Under  the further assumption that the
bundle
${\cal E}$ is trivial, letting $\nabla^{\cal
E}:= d$ be the exterior differential,   the curvature on the
determinant bundle reduces to a trace anomaly:
\begin{eqnarray}\label{eq:Omegares} \left( Det A_x^+\right)^{-1}
\Omega^{Det} Det A^+(U, V) &=& - {1 \over 2}
\partial
{\rm str}^{Q_x }\left(\left(A_x \right)^{-1}  d
   A (U),\left(A_x \right)^{-1}  d A (V)
\right) \nonumber  \\
&+&{1 \over 2} [d{\rm str}^{Q_x }]
\left( \left(A_x \right)^{-1}  dA   \right) (U, V)
\end{eqnarray}
and hence  a trace anomaly residue by   {\rm (\ref{eq:scoboun})} and
{\rm (\ref{eq:dstr})}.
\end{enumerate}
\end{thm}
{\bf Remark.} (\ref{eq:Omegares}) is a
particular case of a more general formula
obtained in \cite{PR}, where no assumption was made
on the triviality of the determinant bundle:
\begin{eqnarray*}  \left(\left( Det A_x \right)^{-1}
\Omega^{Det} Det A \right)(U, V)&=&
-{\rm str}^{Q_x}(\Omega^{\cal E}) (U, V)\\
&-& {1 \over 2}
\partial
{\rm str}^{Q_x}\left(A_x^{-1} [\nabla_U^{
\cal E}, A ] , A_x^{-1} [\nabla_V^{
\cal E}, A ]\right) \\
&+&{1 \over 2} [\nabla^{ \cal E}, {\rm str}^Q]
\left(A_x^{-1} [\nabla ^{
\cal E}, A^+]\right) (U, V) )
\end{eqnarray*}
which yields back (\ref{eq:Omegares}) when taking
$\nabla^{\cal E}:= d$.
The particular case under consideration here of
a trivial determinant bundle is sufficient when
studying gauge anomalies while the more general
setting of \cite{PR} would  be necessary to investigate
gravitational anomalies.
\\

{\it Proof.}
\begin{enumerate}
\item
\begin{eqnarray*}  \left(Det A_x \right)^{-1}
  \nabla^{Det} Det A &-& \left(Det A_x \right)^{-1}
\tilde \nabla^{Det} Det A
  \\ &= &\left[{\rm tr}^{
A_x }- {\rm tr}^{\vert A_x
\vert}\right]
\left(  A_x^{-1}
[\nabla^{ {\cal E}} ,
A ]\right)  \\
&= & -{1 \over a} {\rm res}\left(
A_x^{-1}
[\nabla^{ {\cal E}} ,
A ] \left( \log A_x- \log \vert A_x
\vert\right)\right).
\end{eqnarray*}
This combined with
Lemma \ref{l:BFcreal} yields  (\ref{eq:Nablares}). Differentiating on
either side yields the expression of the
curvature as the differential of a trace anomaly
residue.
\item  A straightforward computation
in the spirit of that of Lemma \ref{l:Retr2} yields:
$$  \left( Det A_x^+\right)^{-1}
\nabla^{Det} Det A^+=d\log {\rm det}_\zeta \vert
A_x^+\vert + {1 \over 2}
{\rm str}^{Q_x}\left( A_x^{-1}
[\nabla^{ \cal E}, A]\right)$$
the weighted supertrace
corresponding to the purely imaginary part of the
connection, the exact form to the real part as
shown in Lemma \ref{l:Retr2}. Here
$Q:=A^2$. Specializing to $\nabla^{\cal E}= d$ in
the case of a trivial bundle ${\cal E}$ and
differentiating this expression yields:
\begin{eqnarray*}  \left(
Det A_x^+\right)^{-1}
\Omega^{Det} Det A^+&=&  {1 \over 2}
d\left({\rm str}^Q\left(A^{-1} d
A \right)\right)\\
&=& {1 \over 2}
[d {\rm str}^Q]\left(A^{-1} d
A \right) -{1 \over 2}
 {\rm str}^Q\left(A^{-1} d
A A^{-1} d
A \right).
\end{eqnarray*}
Formula (\ref{eq:Omegares}) then follows
applying this formula to the vectors $U$ and $V$.
\end{enumerate}
\endsquare
\section{The Chern Simons term  as an
integrated trace anomaly }
In this section and the next one, we specialize  to
the case of a trivial bundle
${\cal E}\to X$. We therefore consider a {\it
trivial} fibration
$\M\to X$, with {\it constant} fibre given by    a
closed   spin  manifold
$M$ and a   Hermitian Clifford vector bundle
$\E\to
\M$ with {\it constant} fibre given by a
     Hermitian Clifford vector bundle
$E= S\otimes W$ where $S$ is the spin bundle and
$W$ an exterior bundle on
$M$. Thus
$${\cal E}\simeq X\times \Ci(M, E).$$
Note that  in the context of gauge theory, $W=
adP$ where $P$ is typically an $SU(N)$ (non abelian
case) or an
$U(1)$ (abelian case) principal bundle on $M$.
\\  We specialize here to the odd dimensional
case, leaving the even dimensional case for the
next section.\\
\\ 
To a smooth family of Hermitian connections
$\{\nabla^W_x, x\in X\}$ on $W$, we associate a
smooth family of Clifford connections
$\{\nabla^{L.C.} \otimes 1+ 1
\otimes
\nabla_x^W, x\in X\}$, where $\nabla^{L.C.}$ is the
  Levi-Civita connection on
$M$ given by a Riemannian metric. These Clifford
connections,  combined with the Clifford
multiplication
$c$, yields a family of  Dirac operators   acting on
smooth sections
$\Ci(M, E)$ of the Clifford module $E$ (see e.g.
\cite{BGV}, \cite{LaMi}, \cite{Fr}):
\begin{equation}\label{eq:Diracx} \{D_x:=
c\circ (\nabla^{L.C.} \otimes 1+ 1
\otimes
\nabla_x^W), x\in X\}.
\end{equation}
Since the
underlying manifold is odd-dimensional they
are formally self-adjoint.
\subsubsection*{The signature operator on a
$3$-dimensional   manifold }
 We  apply  the result of Theorem \ref{th:etatrace} and
its corollary  to the signature operator on an
{\it odd dimensional } manifold $M$.   Let
$\rho$ be a representation of the fundamental
group of
$M$ on an inner product space
$V$ and let $W$ be the vector bundle over
 $M$ defined by $\rho$. The bundle $E:=\oplus_k
\Lambda^k T^*M
\otimes W$ is a Clifford module  for the
following Clifford multiplication:
\begin{eqnarray*} \Ci(T^*M) \times \Ci(E) &\to& \Ci(E) \\
(a, \alpha )&\mapsto& \e(a)
\wedge \alpha -i(a)
\alpha
\end{eqnarray*}
where $\e(a)$ denotes exterior product, $i(a)$ interior product. It can
also be equipped with a
Hermitian structure coming from that on $W$ and the natural inner product
on forms induced by the
Riemannian structure on $M$. The Clifford bundle is naturally graded by
the parity on forms:
$$ E:= E^+ \oplus E^-=\left(\oplus_i\Lambda^{2i} T^*M \otimes W
\right)\oplus \left(
\oplus_i\Lambda^{2i+1} T^*M \otimes W \right).$$
Let $\Omega^k:= \Ci(\Lambda^k T^*M\otimes W)$ be
the space of smooth $W$ valued  $k$-forms on
$M$. We henceforth assume the de Rham complex $0\to
\Omega^0\to \Omega^1\to \cdots \to \Omega^n$ is
acyclic. The bundle
$W$ comes with a flat (self-adjoint) connection
$\nabla^\rho$ that couples with  the Levi-Civita connection $\nabla^{L.C.}$
to give a
(self-adjoint) connection
$\nabla= \nabla^{L.C.} \otimes 1 \oplus 1\otimes \nabla^\rho$ on $E$ from
which we can construct a
Dirac operator
$D_\nabla$. On the other hand, the exterior
differentiation $d$ coupled with   the flat
connection
$\nabla^\rho$ yields an operator
$d_\rho:= d
\otimes 1 + 1 \otimes \nabla^\rho: \Ci(E)
\to \Ci(E) $ such that $d_\rho^2=0$. Identifying
$d$ with
$\e \circ \nabla^{L.C.}$,
$d^*$ identifies to $-i \circ \nabla^{L.C.}$, 
from which it easily follows that $d +d^*=
 (\e- i)
\circ \nabla^{L.C.}= c \circ \nabla^{L.C.}$ and
hence
$$D_\nabla:= c \circ (\nabla^{L.C.} \otimes 1 + 1
\otimes \nabla^\rho)=   d_\rho+ d_\rho^*.
$$ In the following we drop the explicit mention
of the representation $\rho$ in the notation
writing $d$ instead of $d_\rho$ and denoting by
$d_k$ its restriction to $k$ forms.  Note that in
dimension $n=2k+1$, the operator $*d_k$, where $*$
denotes the Hodge star operator, is a formally
self-adjoint elliptic operator of order $1$. We
need to further restrict it in order to get an
invertible operator. The
complex
$0\to \Omega^0\to \Omega^1\to \cdots \to \Omega^n\to
0$ being acyclic, we can write $\Omega^k
= \Omega_k^\prime\oplus \Omega_k^{\prime
\prime}$ where
$\Omega_k^\prime= {\rm Im} d_{k-1}= \ker d_k$ and
$\Omega_k^{\prime \prime}= \ker d_{k-1}^*=
\hbox{ Im
 } d_{k-1}^*$. Restricting the operator $*d_k$ to
$\Omega_k^{\prime \prime}$:
$$*d_k^{\prime \prime} :=
{*d_k}_{\vert_{\Omega_k^{\prime \prime}}}$$ yields
in dimension $n= 2k+1$, an invertible  formally
self-adjoint elliptic operator of order $1$. In
the following proposition, we  first let the
connection $\nabla^W$ vary, then the metric $g$
on $M$ vary, which give rise to two families of
self-adjoint operators to which we shall apply
Corollary \ref{c:phaseres} of section 3.
 \begin{prop}\label{p:detoverdet}
 Let $M$ be a
$3$-dimensional closed Riemannian manifold.
Using the above construction, with $n=3$ ($k=1$), 
one can build a smooth family of self-adjoint
operators
$\{D_t:= *d_{ 1,  t}^{\prime \prime}, t\in [0,
1]\}$ from:
\begin{itemize}
\item a smooth family of connections
$\{\nabla_t^W:= \nabla_t^\rho, t\in [0, 1]\}$ on
$W$ and a fixed metric on $M$
\item  or  a smooth family
of Riemannian metrics $\{g_t, t\in [0, 1]\}$
(inducing a family of Levi-Civita connections)
and a fixed connection
$\nabla^W$ on $W$.
\end{itemize}
In both cases, the  phases
$\phi(D_0), \phi(D_1)$ of the
 $\zeta$-determinants of $D_t$ at the end points $t=0$ and
$t=1$, given by {\rm (\ref{eq:detodd})},
differ by  a Wodzicki residue coming from
 an integrated  trace anomaly:
\begin{eqnarray}\label{eq:detoverdet}
\phi(D_1)- \phi(D_0) & = &  {\pi \over 2} (\eta_{D_1}(0)-
\eta_{D_0}(0))  \nonumber \\
 & = &
  {\pi\over 2}
\int_{0}^{1} {\dot {\rm tr} }(  D_t  \vert
D_t \vert^{-1})dt \nonumber\\
&=&
 -{\pi\over 2}
\int_{0}^{1} {\rm res} ( \dot  D_t  \vert
D_t \vert^{-1})dt
\end{eqnarray}
\end{prop}
{\bf Remark.}  The local
expression on the right hand side corresponds to
the local term given  by the
Atiyah-Patodi-Singer  theorem  \cite{APS II} in terms of
underlying characteristic classes as we shall see
in   Appendix B.
\\

 {\it Proof.} Since  the
signature of
$M\times [0, 1]$ vanishes, so does the spectral
flow of the family $\{D_t, t\in [0, 1]\}$,  so that
the assumptions of Corollary  \ref{c:phaseres} are satisfied.
Applying Corollary  \ref{c:phaseres} yields the result.
\endsquare
\subsubsection*{The Chern Simons model}
Let us give an intepretation of  formula (\ref{eq:detoverdet})
in the context of gauge theory as a {\it phase
anomaly} of some partition function.
\\  
The
Chern-Simons  model
 in dimension $n=2k+1$ (\cite{Sc},   \cite{AdSe}) is
described in terms of a classical action
functional  of the type $S_k(\omega_k)= \langle
\omega_k, *d_k
\omega_k\rangle$, which  presents a degeneracy.
Here $\langle \alpha, \beta\rangle= \int
\alpha \wedge * \beta$ for any $p$-forms
$\alpha$ and $\beta$, where $*$ is the Hodge
star operator. Indeed, writing
$\omega_k= \omega_k^\prime\oplus
\omega_k^{\prime \prime}$ in the above mentioned decomposition, we have
$S_k(\omega_k)=
S_k (\omega_k^{\prime \prime})$. To deal with this type of degeneracy,
A. Schwarz \cite{Sc} suggested
--in analogy with the Faddeev-Popov procedure--
to define the partition function  associated to the classical action
functional  $S_k$  by the
following :
\\ \\
{\bf Ansatz.}
 \begin{eqnarray*}  
Z_k &``:="& \int_{\Omega^j} {\cal D}
\omega_k e^{- \langle \omega_k, *d_k
\omega_k\rangle} \\
& ``:=" & \left( \prod_{l=0}^{k-1} \left({\rm det}_\zeta(\Delta_l^{\prime
\prime}\right)^{(-1)^{k-l+1}} \right)^{1
\over 2}\int_{\Omega_k^{\prime \prime} }{\cal D} \omega_k^{\prime \prime}
e^{- \langle
\omega_k^{\prime \prime}, *d_k \omega_k^{\prime \prime}\rangle }\\
 &  = &
\left( \prod_{l=0}^{k-1}\left( {\rm det}_\zeta(\Delta_l^{\prime
\prime})\right)^{(-1)^{k-l+1}} \right)^{1\over 2}
{\rm det}_\zeta\left(* d_k^{\prime \prime}\right)^{-{1 \over 2}}
\end{eqnarray*}
where we have inserted inverted commas around  identities involving
heuristic objects  such as
${\cal D} \omega_k$, which are to be understood on
a heuristic level. However, the last formula is
well defined since in $n=2k+1$ dimensions  the
operator
$*d_k$ is self-adjoint and hence has a
well-defined determinant.    Using Hodge duality
and the fact that
$\vert {\rm det}_\zeta(* d_l^{ \prime \prime})\vert =
\sqrt {{\rm det}_\zeta(\Delta_l^{\prime
\prime})} $ it follows that:
$$\vert Z_k\vert= \sqrt{   T(M)^{(-1)^{k+1}}}$$
where $T(M)$ is the analytic torsion of
$M$\cite{RS}:
\begin{equation}\label{eq:antor}
T(M):=  \prod_{j=0}^{k}
{\rm det}_\zeta (\Delta_j^{\prime
\prime})^{{(-1)^{j-l+1}}\over 2}.
\end{equation}
Let us comment on the notations used in this
formula, in particular on the meaning of the
$\zeta$-determinants which are involved in the
formula. Restricting the operator
$\Delta_k:=\Delta_{\nabla_{\vert_{\Omega^k}}}=
  d_{k }^*d_k + d_{k-1}
d_{ k-1}^*$ to $\Omega_k^{\prime \prime}$, we get an invertible operator
$\Delta_k^{\prime
\prime}:= { d_{k }^*d_k}_{\vert_{ \Omega_k^{\prime \prime}}}$. As the
restriction to
$\Omega_k^{\prime \prime}$ of a self adjoint elliptic operator, the
operator $\Delta_k^{\prime
\prime}$   has purely discrete real spectrum $\{\l_n^{\prime
\prime}, n \in \N \}$  and the usual
$\zeta$-function techniques can be extended to define
${\rm det}_\zeta(\Delta_k^{\prime \prime}):=
\exp\left(-\zeta_{\Delta_k^{\prime \prime}}^\prime(0)\right)$, where
$\zeta_{\Delta_k^{\prime \prime}} (s) := \sum_{n} \left(\lambda_n^{\prime
\prime}\right)^{-s}$
  see \cite{RS}. \\
\\
 Writing
${\rm det}_\zeta(*d_k^{\prime \prime}) =
\sqrt{{\rm det}_\zeta  \Delta_k^{\prime \prime}} e^{i{\pi \over 2} \left(
\eta_{*d_k^{\prime
\prime}} (0) - \zeta_{*d_k^{\prime \prime}}
 (0)\right) }$ as in formula (\ref{eq:detphase})
  we find:
\begin{equation}\label{eq:Zk}
Z_k=  \sqrt{ T(M)^{(-1)^{k+1}}}
 e^{-i{\pi \over 4}
 \eta_{*d_k^{\prime\prime}} (0)   }
\end{equation}
where we have used the fact that
  $ \zeta_{ \vert *d_l^{\prime \prime} \vert  }(0)=0$  in odd
dimensions. This yields back the fact
that $\vert Z_k\vert=  \sqrt{
T(M)^{(-1)^{k+1}}}$.  \\ \\
 A variation of  the
underlying metric  on
$M$ induces a variation of   the partition
function. The analytic torsion being a
topological invariant,
 its modulus remains constant and  it follows from
Proposition 2 that the phase of the partition
function  changes by some local Wodzicki residue
term. In
\cite{Wi} (see also \cite{At}), Witten   suggested
to modify  this  partition function adding  such
local counterterms in order to build a  regularized
partition function  independent   of the metric on
$M$.   For this he proceeded in two steps,
first fixing the metric and measuring the
dependence of the phase on the choice of
connection and then, whenever the manifold $M$
has trivial tangent bundle,  fixing the connection
and  measuring the
dependence of the phase on the choice of
metric. Both these
dependences can be measured in terms of tracial
anomalies along the lines of Proposition
\ref{p:detoverdet}. Since the classical action
$\langle
\omega_k^{\prime \prime}, *d_k \omega_k^{\prime
\prime}\rangle= \int \omega_k^{\prime
\prime}\wedge d_k
\omega_k^{\prime
\prime} $  is independent of the
choice of the metric, the dependence of the phase
of the partition function  on the metric arises as
an anomaly on the quantum level, which we shall
refer to as a   {\it phase anomaly} of the
partition function.  By Proposition
\ref{p:detoverdet}, the variation of the
partition function
$Z_k(g_0)\to Z_k(g_1)$ induced by a
change of metric
$g_0\to
g_1$  reads:
$$ {Z_k(g_1)\over
Z_k(g_0)}=\exp \left( -i {\pi \over 4}
(\eta_{ *d_{k,1}}(0)-
\eta_{ *d_{k,0}} (0))\right)    $$where as in
Proposition \ref{p:detoverdet}, $\{g_t, t\in [0, 1]\}$ is a
family of Riemannian metrics interpolating $g_0$
and $g_1$, the connection $\nabla^W$ on $W$
being left fixed.    For $k=1$, and when the
tangent bundle is trivial --in which case we can
write the Levi-Civita connection $\nabla^{L.C.}= d+
\omega$-- it  gives rise, via the
Atiyah-Patodi-Singer theorem (see Appendix B), to
the familiar  Chern-Simons  term $ \int_M {\rm tr}
\left (\omega \wedge d\omega  + {2 \over 3}
\omega \wedge \omega \wedge \omega\right)$ arising
in topological quantum field theory in dimension
$3$ (cfr. formula (2.20) in \cite{Wi}). \\
\\  Proposition  \ref{p:detoverdet} thus
establishes a correspondence between:
$$
\begin{array}{ccccc}
{\begin{tabular}{|p{2.9cm}|}
\hline
{\it phase} anomaly 
 for \\ the Chern-Simons partition function \\
\hline
\end{tabular}}
& \Leftrightarrow &
{\begin{tabular}{|p{2.7cm}|}
\hline
 {\it tracial}
anomaly \\  $\int_0^1 \dot {{\rm tr}^{A_t}}({\rm
sgn} A_t) dt$
\\
\hline
\end{tabular}} & \Leftrightarrow &
{\begin{tabular}{|p{3.9cm}|}
\hline
  local term in the \\
Atiyah-Patodi-Singer
{\it index theorem} for families \\
\hline
\end{tabular}}
\end{array}
$$
\section{Chiral (gauge) anomalies}
\subsubsection*{A determinant  bundle on the space
of connections}
 We consider here   an {\it even
dimensional} closed Riemannian manifold $M$  in
which case the spinor bunlde $S$ splits
$S=S^+\oplus S^-$ and the Clifford module
$E= S\otimes W$ splits accordingly into $E=
E^+\oplus E^-$.\\ \\
Let
$X:= {\cal C}(W)$ denote the affine space of
connections on  the exterior bundle $W$ based on
$M$.   ${\cal C}(W)$ is an affine Fr\'echet space
with vector space $\Omega^1(M, Hom(W))$, the
space of $Hom(W)$-valued on forms on $M$.
Concretely, this means that fixing a reference
connection
$\nabla_0^W\in {\cal C} (W)$ (e.g. the ordinary
exterior differentiation if $W$ is trivial), any
other connection  reads
$\nabla^W= \nabla_0^W + A$ where $A$ is a $Hom(W)$
valued one form on $M$. We henceforth use this
reference connection to identify $\nabla_A^W$
with the $1$-form $A$. \\
\\ To the smooth family of   connections
$\{\nabla^W_A, A\in {\cal C}(W)\}$ on $W$, we
associate a smooth family of Clifford connections
$\{\nabla^{L.C.} \otimes 1+ 1
\otimes
\nabla^W_A, A\in {\cal C}(W)\}$, which combined
with the Clifford multiplication $c$ yields a
smooth  family of {\it chiral} Dirac operators
acting from
$\Ci(M, E^+ )$ to $\Ci(M, E^-)$:
\begin{equation}\label{eq:DiracA+} \{D_A^+:=
c\circ (\nabla^{L.C.} \otimes 1+ 1
\otimes
\nabla_A^W), A\in {\cal C}(W)\}.
\end{equation}
Associated to the family
$\{D_A^+, A\in {\cal C}(W)\}$, there is a
determinant bundle  ${\cal L}_{D^+}$ on $W= {\cal
C}(W)$, which is trivial since the manifold $M$ is
kept fixed here.\\ We set as before
$$D_A^-:= \left(D_A^+\right)^*, \quad
\Delta_A^+:= D_A^-D_A^+, \quad \Delta_A^-:= D_A^+
D_A^-, \quad \Delta_A:= \Delta_A^+\oplus
\Delta_A^-.$$
\subsubsection*{The gauge group  action}
The
gauge group
${\cal G}:=
\Ci(M, Aut(W))$  is a Fr\'echet Lie group
with Lie algebra
$Lie\left({\cal G}\right):=
\Ci(M, Hom(W))$. If $W= ad P$ where $P\to M$ is a
trivial principal $G$ bundle, $G$ the structure
group,  then $Lie\left({\cal G}\right):=
\Ci(M, Lie G)$ where $Lie(G)$ is the Lie algebra
of $G$.
\\ The gauge group acts on
${\cal C}(W)$ by:
\begin{eqnarray*} \Theta: {\cal G}\times {\cal C}(W) &\to&
{\cal C}(W)\\
(g, \nabla^W) &\mapsto& g^*\nabla^W
\end{eqnarray*}
and induces
a map:
\begin{eqnarray*} \theta_A: {\cal G}\  &\to&
{\cal C}(W)\\
 g  &\mapsto& g^*\nabla^W.
\end{eqnarray*}
This map is not injective unless  the connection
$A$ is irreducible.   \\ \\
Identifying the tangent space  $T_e{\cal G}$ at the
unit element $e$ of ${\cal G}$ with the
Lie algebra
$Lie\left({\cal G}\right)$, the tangent map  reads:
\begin{eqnarray}\label{eq:tanmap}
d_e\theta_A :  Lie\left({\cal G}\right) &\to&
T_A{\cal C}(W) \nonumber\\
u &\mapsto & {d \over dt}_{\vert_{t=0}} (g_t^*
\nabla_A^W)= [\nabla_A^W, u]
\end{eqnarray}
where $g_t:= {\rm exp} tu$,
 exp  being the exponential map on the gauge group
$\Ci(M, Aut(W))$ (which one might want to
complete into a Hilbert Lie group at this stage
but we shall skip these technicalities here). \\
\\ The BRS (Becchi-Rouet-Stora) operator is
defined by:
\begin{eqnarray*} \delta: \Omega^1({\cal G}  ,
\Omega^1(M, Hom(W)))  &\to& \Omega^2({\cal G}  ,
\Omega^1(M, Hom(W)))
\\
\alpha \otimes A &\mapsto&  d\alpha \otimes A-
\alpha \otimes d \theta_A
\end{eqnarray*}
where $A\in
\Omega^1(M, Hom(W))$.  It is clear from its
definition that
$\delta^2=0$ so that one can define the
corresponding cohomology, called BRS cohomology.
It moreover follows from the above definition
that:
$$ \delta A =
- d \theta_A(\omega), \quad \delta \omega= -{1
\over 2} [\omega, \omega]=-
\omega \wedge \omega= -\omega^2$$
where $\omega $  is the Maurer-Cartan form on
${\cal G}$, namely the left invariant $Lie G$
valued one form on ${\cal G}$  defined by
$\omega_e(v)= v$ for
$v\in Lie \left({\cal G}\right)$. It is called  the
Faddeev-Popov ghost
 and written $\omega= g^{-1} dg$  in the BRS
context.
\subsubsection*{The pull-back of the
Bismut-Freed connection by the gauge group
action}
 Since the line bundle
${\cal L}_{D^+}$ (using the notations of section 4)
is trivial, we can take
$\nabla^{{\cal E}^+}=
\nabla^{{\cal E}^-}= d$ and equip it with the
Bismut-Freed connection $\nabla^{Det}$ defined in
(\ref{eq:sNablalog}) with $d$ instead of $\nabla^{Hom({\cal E}^+,
{\cal E}^-)}$.
 \\ Given a connection $A$,
the Bismut-Freed
 connection  on the line bundle ${\cal
L}_{D^+}\to {\cal C}(W)$    can be pulled back by
the map
$\theta_A$ to a one form on the gauge group
${\cal G}$:
\begin{equation}\label{eq:thetaA}
\left(\theta_A^* \nabla^{Det}\right)_u=
\nabla^{Det}_{\bar U_A}
\end{equation}
where $\bar U_A:= d\theta_Au$ is  the canonical
vector field on ${\cal C}(W)$  generated by $u\in
Lie\left({\cal G}\right)$.
The following proposition
expresses the pull-back of the Bismut-Freed
connection in  the direction of 	 $u\in Lie
\left({\cal G}\right)$:
\begin{prop}\label{P:thetaNabla} Given an
irreducible connection $A$, the pull-back
$\theta_A^* \nabla^{Det}$ of the Bismut-Freed
connection on the gauge group in the direction
$u\in Lie\left({\cal G}\right)$  is a
local expression which can be interpreted as a
chiral gauge anomaly. Given a section $Det D^+_A$
of ${\cal L}_{D_A^+}$  which is invertible at
$A$:
\begin{equation}\label{eq:thetaNabla}
\left( Det
D_A^+ \right)^{-1} \left(\theta_A^*
\nabla^{Det}\right)_u Det D_A^+ =
 {\rm str}^{\Delta_A } (u)=
\int_M \hat A(\nabla^{L.C.})
{\rm tr}_m \left( e^{-\Omega_A^W}
u(m)\right)d\mu(m)
\end{equation}
 where $\Omega_A^W$ is the curvature of
$\nabla_A^W$, ${\rm tr}_m$ the trace on the
fibre $W_m$ above $m$ and $\hat A(\nabla^{L.C.})$
the $\hat A$-genus on $M$.
\end{prop}

{\it Proof.} It follows from
definition (\ref{eq:sNablalog}) that:
\begin{eqnarray*} \left(Det
D_A^+\right)^{-1}\left(\theta_A^*
\nabla^{Det}\right)_u Det D_A^+  &=&
{\rm tr}^{\vert D_A^+\vert} \left(\left(
D_A^+\right)^{-1}  (dD_A^+)(\bar U_A)\right)\\
&=& {\rm tr}^{\vert D_A^+\vert} \left(\left(
D_A^+\right)^{-1}c\circ (d \nabla_A^W(\bar U_A))
 \right)\\  & =&
{\rm tr}^{\vert D_A^+\vert} \left(\left(
D_A^+\right)^{-1}   \left(c\circ[\nabla_A^W,u]
\right)  \right)\\
& =&
{\rm tr}^{\vert D_A^+\vert} \left(\left(
D_A^+\right)^{-1}  [ D_A^+ , u]\right) \\
& =&
{\rm tr}^{\vert D_A^+\vert} \left(\left(
D_A^+\right)^{-1}    D_A^+   u \right)-
 {\rm tr}^{\vert D_A^+\vert} \left(\left(
D_A^+\right)^{-1}  u  D_A^+     \right)
\\
 & =&
{\rm tr}^{\vert D_A^+\vert} \left(   u \right)-
 {\rm tr}^{\Delta_A^+ } \left(\left(
D_A^+\right)^{-1}  u  D_A^+     \right) \\
 & =&
{\rm tr}^{\vert D_A^+\vert} \left(   u \right)-
 {\rm tr}^{\Delta_A^- } \left(D_A^+  \left(
D_A^+\right)^{-1}  u     \right) \\
 & =&
{\rm tr}^{\Delta_A^+} \left(   u \right)-
 {\rm tr}^{\Delta_A^- } \left(   u     \right) \\
&=& {\rm str}^{\Delta_A}(u)
\end{eqnarray*}
where we have
used the fact that $D_A^- \Delta_A^+= \Delta_A^+
D_A^-$ as can  easily be checked from  the
definition  of $\Delta_A^+$.  This proves the first
equality in (\ref{eq:thetaNabla}).  The local version of the
Atiyah-Singer theorem then yields a  local
expression for the   term
${\rm str}^{\Delta_A }(u
 ) $. Indeed   it follows from results by Patodi
and Gilkey  that (see e.g. Theorem 4.1 in \cite{BGV})
$$   k_\e(m, m) \sim (4 \pi t)^{- {n\over 2}}
\sum_{i=0}^\infty t^i k_i (m)$$
$k_\e(m, n)$,
$m, n\in M$ is the kernel of the heat-operator
$e^{-\e
\Delta} $ and $k_i \in \Ci(M, C_{2i}(T^*M) \otimes Hom(W))$. Thus, applying
fibrewise $u(m)$, taking the trace on the fibre above $m$  and
then integrating along $m$ we get:
\begin{equation}\label{eq:intfibre}
\int_M  {\rm str}_{m} (u(m)  k_\e(m, m)) \sim (4
\pi t)^{- {n\over 2}} \sum_{i=0}^\infty t^i
\int_M  {\rm str}_{m}(u(m) k_i (m))
\end{equation}
where ${\rm str}_{m}$ means we have taken the supertrace
along the ($\Z_2$-graded) fibre $E_m$ of $E$ above
$m\in M$. On the other hand, the pointwise
supertrace
 ${\rm str}_{m} (a \otimes b)$ of $a\otimes b \in
C (T_m^*M)\otimes Hom(W_m)$ is equal to a Berezin
integral (see e.g. Prop 3.21 in \cite{BGV}):
$$  {\rm str}_{E_m} (a \otimes b) = (-2i)^{n \over
2}\sigma_n(a(m)) {\rm str}_{W_m}  b(m)$$
where $\sigma$
is the symbol map taking Clifford elements to
forms. Combining this with  (\ref{eq:intfibre}) eventually yields
the local  expression $ {\rm str}^{\Delta_A
}(u ) = \int_M \hat A(\nabla^{L.C.})
{\rm tr}_m \left[ \Omega_A^W
u(m)\right]d\mu(m)$ after making the usual
identifications with the underlying geometric data. \endsquare
\\ \\
 The pull-back
$\theta_A^* \nabla^{Det}$ on the gauge
group measures a chiral gauge anomaly; there is an
apriori obstruction for it to be consistent,
namely the pull-back of the curvature $\theta_A^*
\Omega^{Det}$ of the Bismut-Freed connection,
which measures the obstruction to the
{\it Wess-Zumino consistency relations} for this gauge
anomaly.
\begin{prop}\label{p:WZobs} The
obstruction to the {\rm  Wess-Zumino consistency 
relations} for the gauge anomaly given by the
pull-back
$\theta_A^* \Omega^{Det}$  of
the Bismut-Freed connection on the gauge group  is
measured by the pull-back of its  curvature.  It has a local feature 
since it is a Wodzicki residue arising from trace anomalies. Using the 
index theorem for families, it can be expressed as an integral on $M$ 
of some local form:
$$ \left( Det D_A^+\right)^{-1}
\theta_A^*\Omega^{Det} Det D_A^+(u, v)= \lim_{\e
\to 0} \int_M \hat A(\Omega^{L.C.})
{\rm str}_x\left(e^{-(\sqrt \e D_A+ \e [d,
D_A])^2}
\right)_{[2]}(\bar U_A, \bar V_A)$$
 where $\bar U_A:= d\theta_Au, \bar V_A=
d\theta_A v, u, v\in Lie\left({\cal G}\right)$   and
$\sqrt \e D_A+ \e [d, D_A]$ the part of degree $1$ of the
family parametrized by $\e$ of  superconnections
associated to the family
$D_A, A\in {\cal C}(W)$  {\rm \cite{Q2}, \cite{BF}, \cite{BGV}}.
\end{prop}

{\it Proof.}
   The curvature  of the
Bismut-Freed connection described in formula (\ref{eq:Omegares})
reads:
\begin{eqnarray*}
  \left( Det D_A^+\right)^{-1}
\Omega^{Det} Det D_A^+(U, V) = &-& {1 \over 2}
\partial {\rm str}^{\Delta_A }\left(D_A^{-1}
 dD_A (U), D_A^{-1}  d D_A (V)
\right) \\
& + &{1 \over 2} [d{\rm str}^{\Delta_A}]
\left(D_A^{-1}   dD_A \right) (U, V)
\end{eqnarray*}
which we saw was a combination of trace
anomalies; applying this to  $\bar U_A:= d\theta_Au, \bar V_A=
d\theta_A v, u, v\in Lie\left({\cal G}\right)$ yields the
fact that its pull-back can also be interpreted
as a combination  of trace anomalies and  can
therefore be expressed in terms of Wodzicki
residues using the results of section 1. The
computation of the curvature $\Omega^{Det}$
carried out in \cite{AS} for Dirac opertors
parametrized by connections and later in \cite{BF} in
the case of Dirac operators parametrized by
metrics yields (taking $\nabla^{\cal E}= d$ with
the notations of section 4):
$$ \left( Det D_A^+\right)^{-1}
 \Omega^{Det} Det D_A^+(U,V)= \lim_{\e
\to 0} \int_M
{\rm str}_x\left(\hat A(\Omega^{L.C.}) e^{-(\sqrt
\e D_A+ \e [d, D_A])^2}
\right)_{[2]}( U ,   V )$$ thus leading to the
second part of the proposition. \endsquare
\\ \\
A similar result would
hold for gravitational chiral anomalies
described in \cite{BF} as the curvature on a
determinant bundle associated to a family of
Dirac oeprators parametrized by metrics. The
essential difference is that the geometric
setting there involves a family of Riemannian
(spin) manifolds and   the determinant bundle
associated to the family of Dirac operators is not
trivial. As a result, the curvature on
the determinant bundle is a combination of a
local term given by some trace anomalies and a
local term arising from the underlying geometry of
the fibration fo manifolds; the tracial anomaly
mixes with the underlying geometry to build a
chiral anomaly.
\\ \\ {\bf Concluding Remark.}
This last result shows once again how closely
related (chiral) quantum anoamlies and tracial
anomalies are, thus leading to the
following correspondance scheme:

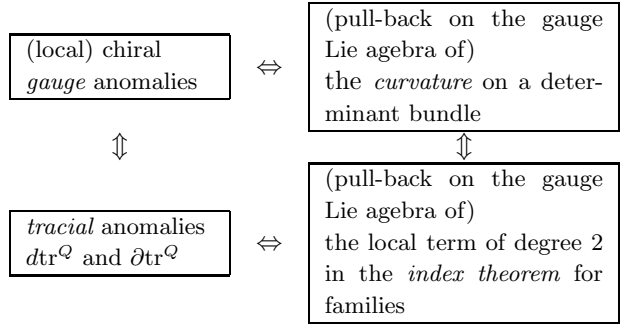
$$
\begin{array}{ccccccc}
{\begin{tabular}{|p{2.5cm}|}
\hline
{\small(local) chiral } \\
{\small  {\it gauge} anomalies} \\
\hline
\end{tabular}}  & \Leftrightarrow &
{\begin{tabular}{|p{3.7cm}|}
\hline
{\small (pull-back on the gauge Lie  agebra of)} \\ the
{\small {\it curvature} on a determinant bundle} \\
\hline
\end{tabular}} \\ \Updownarrow & & \Updownarrow \\
{\begin{tabular}{|p{2.5cm}|}
\hline
{\small {\it tracial} anomalies } \\
{\small $d{\rm tr}^Q$ and $\partial {\rm tr}^Q$} \\
\hline
\end{tabular}} & \Leftrightarrow &
{\begin{tabular}{|p{3.7cm}|}
\hline
{\small (pull-back on the gauge Lie
agebra of)} \\  {\small the local term of degree 2 in the
{\it index theorem} for families }\\
\hline
\end{tabular}}
\end{array}
$$
\newpage
\centerline{\Large {\bf Appendix}}
\appendix
\section{The multiplicative anomaly for
$\zeta$-determinants and anomalies in physics}
In finite dimensions, determinants naturally
arise from Gaussian integration:
$${1 \over (2\pi)^{n\over 2}}\int_{\R^n}
e^{-{1\over 2} <Qx, x>}
d  x  =  (\det Q )^{-{1\over 2}}$$ where
$Q$  positive definite symmetric matrix, $\langle
\cdot,
\cdot
\rangle$ the euclidean inner product on $\R^n$.
   Mimicking the finite dimensional
setting, one computes Gaussian
integrals in infinite dimensions  substituting to
the ordinary determinant, the
$\zeta$-determinant:$$ \int_{{\rm configurations}
\, \varphi } e^{-{1\over 2} < Q \varphi ,
\varphi >} {\cal D}^Q[ \varphi] = ( {\det}_\zeta Q
)^{-{1\over 2}},\eqno(A.1)$$ where
$Q$  is an invertible admissible elliptic
operator with positive order. The integrals on
the infinite dimensional configuration space
of the physical system are therefore to be
understood as the r.h.s. well-defined
$\zeta$-determinant. The ``volume measures" ${\cal
D}^Q[\varphi]
$-- which are there to remind us that we are
mimicking the finite dimensional integration
procedure-- can a priori depend on
$Q$, a dependence one needs to take into account
in the following.
\\ \\
Just as the operator $Q$  ``weights" a priori
divergent traces in a way that enables us extract
a finite part,  it serves here to ``extract
 a finite part" of   a priori ill-defined formal
path integrals.
\\  Let us see how  this $Q$-dependence can
affect the computations.
Starting from the finite dimensional setting, let
us make the change of variable
$\tilde x = Cx$ in a gaussian integral and denote
by
$J$ the corresponding jacobian determinant:
\begin{eqnarray*} (\det Q)^{-{1\over2}}
&=& \int_{\R^n} e^{-{1\over 2}< Q\tilde x,\tilde x
>} d \tilde x\\
&=& \int_{\R^n} e^{-{1\over 2}
<Q Cx , Cx >}J dx\\
& =& J \cdot  \det ( C^* QC)^{-{1\over
2}}.
\end{eqnarray*}
  Furthermore $$J := {
\left(\det (C^*Q C)\right)^{ {1\over 2}}\over (\det
Q)^{ {1\over 2}}
 }=\sqrt{ \det
(C^*C)}= \vert \det
C \vert .$$
Similarly, replacing ordinary determinants
by $\zeta$-determinants, one could expect
the modulus of the jacobian determinant of a
 $\tilde \varphi = C \varphi $ in (A.1)  to correspond to a quotient
of $\zeta$-determinants. But at this point the
multiplicative anomaly comes into the way. \\ \\
Let
$C$ be an invertible elliptic operator (with
possibly zero order),
$C^*$ its formal adjoint (with respect to an $L^2$
structure on the space of sections it is acting
on),assuming that
$Q$ is positive (or ``sufficiently close" to a
positive operator \cite{KV}, \cite{Du}), then
$C^*QC$ is a positive elliptic operator (or
``sufficiently close "  to a positive operator)
with positive order in such a way that we can
define its $\zeta$-determinant. Applying a
computation similar to the finite dimensional
one would yield:
$$  J_Q:= {\det_\zeta (C^*Q C)^{ {1\over
2}}\over (\det_\zeta
Q)^{ {1\over 2}}  }. $$
 But this does not generally coincide  with $$
 \tilde J :=
 \sqrt{ {\det}_\zeta
(C^*C)}.$$In any case the   latter determinant
is only  defined if $C$ has non vanishing
positive order, which is not always the case in
applications where $C$ could typically be a
multiplication operator.  The fact that $J\neq
J_Q$ is a consequence of the multiplicative
anomaly for $\zeta$-determinants recalled in
(\ref{eq:MultAn}) as the following computation shows:
$$   J_Q^2=
{\det_\zeta (C^*Q C) \over \det_\zeta
Q    } =  {
\det_\zeta (Q  C^*C) \over  \det_\zeta
Q  }  = F_\zeta (Q,  C^*C){\det}_\zeta(C^*C)=
F_\zeta (Q,  C^*C)\cdot\tilde J^2.\eqno (A.2)$$
The second identity follows from
interpolating $C^*QC$ and  $QC^*C$  by the family
$Q_t:= Q^tC^*Q^{1-t} C$, $t\in [0, 1]$
of constant order elliptic operators  which have
a constant determinant since:
${d\over dt}
\log \det_\zeta Q_t= 0$. The third identity
folows from (\ref{eq:MultAn}).
\section{Computation of the Chern-Simons term in
TQFT in dimension $3$ using the
Atiyah-Patodi-Singer Theorem } {\bf Theorem
\cite{APS II}} {\it Let  $X$ be an oriented
Riemannian  manifold of dimension $4l$ with
boundary $M$ such that $X$  is isometric to a
product $M\times I, I\subset \R$ near the
boundary. Let
$\nabla^W$ be a connection on the exterior bundle
$W$ based on $X$ and $\nabla^{L.C.}$ the Levi-Civita connection on $X$. Let
$D_\nabla:= d_\nabla+ d_\nabla^*$ where $d_\nabla=  d\otimes 1 +
1\otimes\nabla^W $ and
$d_\nabla^*=  d^*
\otimes 1+ 1
\otimes
\nabla^W $
  as in section 5, and let
$D_\nabla^+$ denote the restriction of
$D_\nabla$ to the even forms on $X$. Near
the boundary, $$D_\nabla^+=
c\circ ({d \over dt}+ B^{odd})
$$ where $B^{odd}$ is the restriction to odd forms on the boundary of the
operator defined on $2p$
or
$2p+1$ forms by: $$B_\nabla= (-1)^{k+p+1} (\e *d_\nabla-d_\nabla*) ,$$
 $\e$ denoting the grading operator on forms. We
let the operator $D_\nabla^+$ act on sections
$f$ of the vector bundle satisfying the
Atiyah-Patodi-Singer (APS) boundary condition
$Pf(\cdot, 0)=0$ where $P$ is the spectral
projection of $B^{odd}$ corresponding to non
negative  eigenvalues. \s Then
$${\rm ind} D_\nabla^+  =
 \int_X L(\nabla^{L.C}){\rm tr}_x\left(e^{
-\Omega^W}\right) +
\eta_{B } (0)  $$  where  $L$ is the
Hirzebruch
$L$ polynomial, $\Omega^W$ the curvature on $W$,
and  where
$\eta_{B }$ denotes the $\eta$ invariant of  $B_{odd}$. }
\\ \\
Let us apply this result   to
$X= M
\times [0, 1]$ where $M$ is an $4l-1$ dimensional closed Riemannian
manifold and let us equip
$X$ with the product metric.
The boundary of
$X$ is  the odd dimensional manifold $  M\times \{0\}
\bigcup M
\times \{1\}$. With the notations of the above theorem where we set $p=k$,
since $k$ is
odd,  we have
$B_k = *d_k- d_{n-k}*$ where $B_k $ is the restriction of $B$ to the odd
$k$ forms. Since  $*^2=
1$ on $k$ forms in dimension $n=2k+1$,   we have
$d_{n-k}^*=- *d_k^*$ so that  the restriction $B_k^{\prime \prime}$ to $R(
d_{k-1}^*)
$ coincides with the restriction  $*d_k^{\prime \prime}$.
\\ In order to compute the r.h.s of (\ref{eq:detoverdet}) we need
to compute the difference of
$\eta$-invariants of $B_k^{\prime \prime}$.
Following Atiyah, Patodi and Singer, let us first
investigate the metric dependence of the eta
invariants $\eta_{*d_k^{\prime \prime}}(0)$ in
order to   build  an invariant independent on the
choice of metric.\s
 \\ To two metrics $g$ and $g^\prime$ on
$M$ correspond two operators
$B   $ and
$B^{ \prime} $ and it follows from the Atiyah-Patodi-Singer index theorem
that (see (2.3) in
\cite{APS II}):
\begin{equation}\label{eq:appB1}
 \eta_{B}(0)-
\eta_{B^\prime}(0)=  n\int_{M \times [0, 1]} L(\nabla^{L.C})
\end{equation}
using the fact
that  sign$(M \times [0, 1])= 0$ and that the
connection on $W$ is flat. \\ Let us now fix the
metric and take  two flat connections
$\nabla_0^W$ and
$\nabla_1^W$ on
$W$ restricted to $ M$, this leading again to two $\eta$ invariants
$\eta_{B_{k,1}^{\prime
\prime}}(0)$ and
$\eta_{B_{k,0}^{\prime
\prime}}(0)$. From the above it follows that this expression is independent
of the choice of metric (see Theorem 2.4 in \cite{APS
II}). \\  We now equip $W$ restricted to $M$ with
  a one parameter family of   connections $\nabla_t^W:= (1-t) \nabla_0^W+ t
\nabla_1^W$ and
correspondingly a one parameter family of operators:
 $$B_t= (-1)^{k+p+1} (\e *d_t-d_t*).$$
   We can equip $W$ seen as a bundle over  $X= [0, 1] \times M$   with the
connection $\nabla^W:= {d
\over dt}+
\nabla_t^W$ and build the corresponding Dirac operator:
$$D_\nabla^+= c\circ ({d \over dt}+ B_t^{odd}).
$$ Because $ B_{k,1}^{\prime
\prime}(0) -B_{k,0}^{\prime
\prime}(0)$ does not depend on the choice of
metric, we can choose a flat metric. Thus the $L$ form will be trivial. On
the other
hand ${\rm sgn}(X)=0$ for the particular choice of manifold $X= M \times [0,
1]$ we took so that the spectral flow vanishes. Applying once again the
Atiyah-Patodi-Singer theorem
yields:
\begin{equation}\label{eq:appB2}
\eta_{B_{k, 1}^{\prime \prime}}(0)-
\eta_{B_{k,0}^{\prime \prime}}(0)=
\int_{M
\times [0, 1]} \tr_x\left(e^{
-\Omega^W}\right) .
 \end{equation}
Combining (\ref{eq:appB1}) and  (\ref{eq:appB2})
where  the Levi-Civita connection reads
$d+ \omega$ and the connection on $W$ reads
$\nabla^W= d+A$ (provided both the tangent bundle
and the bundle $E$ are trivial)  yields the
expression of the  Chern-Simons term computed by
Witten (see formula (2.23) in
\cite{Wi}).
$$ $$
\\
 \\

{\bf Acknowlegments.}
The last author  would like to thank
 Sergio Albeverio, Edwin Langmann and Jouko Mickelsson for giving her the
opportunity to present
 the results of this article at an early stage of their development and for
the fruitful
discussions that followed. She is also very
grateful to Daniel Bennequin who made very useful
comments at a later stage of the  development of
the results presented here.  The authors also
benefitted from useful and interesting discussions
with Jose Gracia Bondia, Matthias Lesch  and  Jean
Orloff  whom we would like to thank most warmly. \\
The first author  was supported by an ECOS-Nord grant during a stay at
the Universit\'e Blaise
Pascal in Clermont-Ferrand where this article was written.

\end{document}